\documentclass[manuscript,screen]{acmart}

\usepackage[utf8]{inputenc}
\usepackage{comment}
\usepackage{titlesec}
\usepackage{array}
\usepackage[most]{tcolorbox}
\PassOptionsToPackage{table}{xcolor}
\usepackage{xcolor}
\usepackage{geometry}
\usepackage{pifont}
\usepackage{tabularx}
\usepackage{tikz}
\usepackage{textcomp}
\usepackage{makecell}
\usepackage{xspace}
\usepackage{amsmath}
\usepackage{amssymb}
\usepackage{enumitem}
\usepackage{booktabs}
\usepackage{ragged2e}
\usepackage{rotating}
\usepackage{fontawesome5}
\usepackage{caption}
\usepackage{colortbl}
\usepackage{multirow}
\usepackage{graphicx}
\usepackage{mdframed}
\usepackage{needspace}

\tcbuselibrary{skins, breakable}

\definecolor{lightgray}{rgb}{0.95, 0.95, 0.95} 
\definecolor{darkgray}{rgb}{0.7, 0.7, 0.7} 
\definecolor{mygreen}{rgb}{0.0, 0.5, 0.0} 
\definecolor{myred}{rgb}{0.8, 0.0, 0.0} 

\newcommand*\circled[1]{\tikz[baseline=(char.base)]{
            \node[shape=circle,fill,inner sep=0.8pt] (char) {\textcolor{white}{#1}};}}

\newcolumntype{C}{>{\bfseries}c} 
\definecolor{main}{HTML}{cccccc}    
\definecolor{sub}{HTML}{000000}     

\newcommand{\circledletter}[1]{%
  \tikz[baseline=(char.base)]\node[shape=circle,draw=black,fill=black,inner sep=1pt] (char) {\textcolor{white}{\textbf{#1}}};%
}

\newtcolorbox{boxM}{
    fontupper = \color{blue},
    rounded corners,
    arc = 6pt,
    colback = main!80, 
    colframe = main, 
    boxrule = 0pt, 
    bottomrule = 4.5pt,
    enhanced,
    fuzzy shadow = {0pt}{-3pt}{-0.5pt}{0.5pt}{blue!35},
    overlay={ 
        \node[anchor=north east] at (frame.north east) {
            \includegraphics[width=15pt]{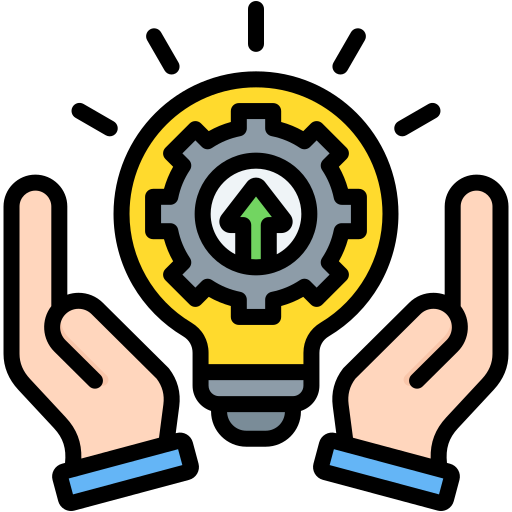}
        };
    }
}

\AtBeginDocument{%
  }

\newcommand{\perfup}{%
\raisebox{-0.35em}{%
\begin{tikzpicture}[scale=0.15]
\draw[thick, black, fill=green] (0,0) circle (1);
\node at (0,0) {\textbf{\textcolor{black}{P}}};
\end{tikzpicture}%
}%
}

\newcommand{\perfdown}{%
\raisebox{-0.35em}{%
\begin{tikzpicture}[scale=0.15]
\draw[thick, black, fill=red] (0,0) circle (1);
\node at (0,0) {\textbf{\textcolor{white}{P}}};
\end{tikzpicture}%
}%
}

\newcommand{\perfneutral}{%
\raisebox{-0.35em}{%
\begin{tikzpicture}[scale=0.15]
\draw[thick, black, fill=blue] (0,0) circle (1);
\node at (0,0) {\textbf{\textcolor{white}{P}}};
\end{tikzpicture}%
}%
}

\newcommand{\effup}{%
\raisebox{-0.35em}{%
\begin{tikzpicture}[scale=0.15]
\draw[thick, black, fill=green] (0,0) circle (1);
\node at (0,0) {\textbf{\textcolor{black}{E}}};
\end{tikzpicture}%
}%
}

\newcommand{\effdown}{%
\raisebox{-0.35em}{%
\begin{tikzpicture}[scale=0.15]
\draw[thick, black, fill=red] (0,0) circle (1);
\node at (0,0) {\textbf{\textcolor{white}{E}}};
\end{tikzpicture}%
}%
}

\newcommand{\effneutral}{%
\raisebox{-0.35em}{%
\begin{tikzpicture}[scale=0.15]

\draw[thick, black, fill=blue] (0,0) circle (1);
\node at (0,0) {\textbf{\textcolor{white}{E}}};
\end{tikzpicture}%
}%
}

\begin{document}

\title{A Systematic Literature Review of Parameter-Efficient Fine-Tuning for Large Code Models}


\author{Saima Afrin$^{*}$}
\affiliation{%
  \institution{William \& Mary}
  \city{Williamsburg}
  \state{Virginia}
  \country{USA}}
\email{safrin@wm.edu}

\author{Md Zahidul Haque$^{*}$}
\affiliation{%
  \institution{William \& Mary}
  \city{Williamsburg}
  \state{Virginia}
  \country{USA}}
\email{mhaque@wm.edu}

\author{Antonio Mastropaolo}
\affiliation{%
  \institution{William \& Mary}
  \city{Williamsburg}
  \state{Virginia}
  \country{USA}}
\email{amastropaolo@wm.edu}

\renewcommand{\shortauthors}{Afrin, Haque, and Mastropaolo}

\thanks{$^{*}$These authors contributed equally to this work.}

\definecolor{darkgreen}{rgb}{0.0, 0.5, 0.0}
\newcommand{\re}{\textcolor{red}{\textbf{[REF]}\xspace}}
\newcommand{\cmark}{\ding{51}}%
\newcommand{\xmark}{\ding{55}}
\newcommand{\ie}{\emph{i.e.,}\xspace}
\newcommand{\eg}{\emph{e.g.,}\xspace}
\newcommand{\etc}{etc.\xspace}
\newcommand{\etal}{\emph{et~al.}\xspace}
\newcommand{\secref}[1]{Section~\ref{#1}\xspace}
\newcommand{\chapref}[1]{Chapter~\ref{#1}\xspace}
\newcommand{\appref}[1]{Appendix~\ref{#1}\xspace}
\newcommand{\figref}[1]{Fig.~\ref{#1}\xspace}
\newcommand{\listref}[1]{Listing~\ref{#1}\xspace}
\newcommand{\tabref}[1]{Table~\ref{#1}\xspace}
\newcommand{\greenAI}{\emph{\textcolor{ForestGreen}{Green} AI}\xspace}
\newcommand{\cTc}{\emph{Code2Code}\xspace}
\newcommand{\cTn}{\emph{Code2NL}\xspace}
\newcommand{\nTc}{\emph{NL2Code}\xspace}
\newboolean{showcomments}
\setboolean{showcomments}{true}

\ifthenelse{\boolean{showcomments}}
{\newcommand{\nb}[2]{
		\fbox{\bfseries\sffamily\scriptsize#1}
		{\sf\small$\blacktriangleright$\textit{#2}$\blacktriangleleft$}
	}
	\newcommand{\cvsversion}{\emph{\scriptsize$-$Id: macro.tex,v 1.9 2005/12/09 22:38:33 giulio Exp $}}
}
{\newcommand{\nb}[2]{}
	\newcommand{\cvsversion}{}
}




\newmdenv[
  linecolor=black,
  linewidth=1pt,
  roundcorner=4pt,
  backgroundcolor=gray!10,
  innerleftmargin=10pt,
  innerrightmargin=10pt,
  innertopmargin=4pt,
  innerbottommargin=8pt,
  skipabove=6pt,
  skipbelow=6pt,
  shadowsize=3pt,
  shadowcolor=gray!20,
  middlelinecolor=black,
  leftmargin=0pt,
  rightmargin=0pt,
  align=center
]{iconbox}

\newcommand{\boxwithsideicon}[3]{%
  \begin{iconbox}
    \begin{minipage}[t]{\linewidth}
      \textbf{#2}%
      \hspace{0.5em}%
      \raisebox{-0.4\height}{\includegraphics[width=0.75cm]{#1}} \\[0.5ex]
      #3
    \end{minipage}%
  \end{iconbox}
}



\newtcolorbox{simpleacademicbox}{
  colback=gray!10,       
  colframe=black!30,
  boxrule=0.3pt,
  arc=1pt,
  left=6pt,
  right=6pt,
  top=4pt,
  bottom=4pt,
}


\newcommand{\verylightgray}[1]{\cellcolor{gray!10}{#1}}
\newcommand{\lightgray}[1]{\cellcolor{gray!22}{#1}}
\newcommand{\gray}[1]{\cellcolor{gray!33}{#1}}
\newcommand{\darkgray}[1]{\cellcolor{gray!45}{#1}}

\newcommand\ALVI[1]{\textcolor{purple}{\nb{ALVI}{#1}}}
\newcommand\ANTONIO[1]{\textcolor{red}{\nb{ANTONIO}{#1}}}
\newcommand\SAIMA[1]{\textcolor{blue}{\nb{SAIMA}{#1}}}

\begin{abstract}
The rise of Artificial Intelligence (AI)-and particularly Large Language Models (LLMs) for code--has reshaped Software Engineering (SE) by enabling the automation of tasks such as code generation, bug detection, and repair. However, these models require significant computational resources for training and fine-tuning, posing challenges for real-world adoption in resource-constrained environments. To address this, the research community has increasingly turned to Parameter-Efficient Fine-Tuning (PEFT)--a class of techniques that enables the adaptation of large models by updating only a small subset of parameters, rather than the entire model. In this Systematic Literature Review (SLR), we examine the growing application of PEFT techniques--across a wide range of software engineering tasks. We analyze how these methods are used to optimize various deep learning (DL) architectures, focusing on their impact on both performance and efficiency. Our study synthesizes findings from 28 peer-reviewed papers, identifying patterns in configuration strategies and adaptation trade-offs. The outcome of this review is a comprehensive taxonomy that categorizes PEFT usage by task type, distinguishing between generative (\eg Code Summarization) and non-generative (\eg Code Clone Detection) scenarios.
Our findings aim to inform future research and guide the practical deployment of PEFT in sustainable, AI-powered software development. \\
\noindent\emph{Our artifacts are publicly available at \url{https://github.com/alvi75/SLR-PEFT}}\end{abstract}



\maketitle


\section{Introduction}
\label{sec:intro}
The field of Software Engineering (SE) has evolved rapidly, driven by the increasing integration of Artificial Intelligence (AI) into core development practices. This shift has introduced new paradigms of automation, fundamentally reshaping how developers approach complex tasks across the software lifecycle.

At the center of this transformation are Large Language Models (LLMs)—and more specifically, Large Code Models (LCMs). These, are deep learning (DL) models that pushed the boundaries of software engineering automation--taming the intricacies of tasks that were once thought to require exclusively human expertise. Activities such as Bug Repair~\cite{zhang2022repairing,joshi2023repair,ahmed2022synshine,fan2023automated,xia2023automated,pearce2023examining,jin2023inferfix,wei2023copiloting,kong2024contrastrepair,weng2023automatic,fu2022vulrepair}, Code Summarization and Documentation~\cite{ahmed2024automatic,virk2024enhancing,yun2024project,ahmed2022few,yun2024project,mastropaolo:icsme2021,mastropaolo2024towards}, Code Completion~\cite{liu2020multi,austin:arxiv2021,guo2023longcoder,izadi2022codefill,ciniselli:tse2021,ciniselli2021empirical}, Software Testing~\cite{mastropaolo2022using,mastropaolo:icse2021,tufano2020unit,tufano:ast2022,schafer2023empirical,ryan2024code}, and Code Review~\cite{tufano:icse2022,lu2023llama,li2022automating,li2022auger} have increasingly benefited from the integration of LCMs, which are capable of learning rich code semantics and generating context-aware predictions. These models not only streamline traditionally manual and error-prone tasks but also demonstrate strong potential for improving productivity \cite{tabarsi2025llms, li2023sheetcopilot} and software quality \cite{widyasari2024beyond, wadhwa2024core} across a wide range of development scenarios \cite{mastropaolo2023towards,mastropaolo2024toward,lubos2024leveraging,yang2024exploring,vitale2023using,mastropaolo2024log,watson2022systematic,hou2024large}.

However, the remarkable performance exhibited by LCMs comes at a significant cost. These models often contain hundreds of millions to billions of parameters, making them exceptionally resource-intensive to train, fine-tune, and deploy. Their sheer scale demands substantial computational power and memory, which in turn raises critical concerns regarding accessibility, scalability, and environmental sustainability—particularly in terms of energy consumption and carbon footprint~\cite{shi2024efficient}. For many organizations and researchers, fully fine-tuning an LCM for each new task is prohibitively costly, creating a pressing need for more efficient adaptation strategies.

To overcome the limitations of full-model fine-tuning, the research community has increasingly embraced Parameter-Efficient Fine-Tuning (PEFT) as a more sustainable and scalable alternative. PEFT techniques are designed to enable model adaptation by modifying only a small portion of a model’s parameters or introducing lightweight, trainable components. Notable approaches in this category include Low-Rank Adaptation (LoRA) \cite{hu2022lora}, Adapter Tuning \cite{wang2022adamix}, and Prompt Tuning \cite{lester2021power}. Rather than updating the entire set of parameters in a large code model, these methods introduce targeted, task-specific modifications—such as low-rank transformations or adapter modules—that significantly reduce both computational demands and GPU memory usage. These strategies have demonstrated performance on par with, or in some cases superior to, full fine-tuning \cite{afrin2025resource, hu2022lora, wang2022adamix, lester2021power, li2021prefix, dettmers2023qlora, zhang2023adalora, pfeiffer2020adapterfusion}, all while requiring only a fraction of the resources. 

In the domain of software engineering, PEFT opens the door to continuous and cost-effective model specialization. For instance, tailoring models to specific repositories, coding styles, or project needs--without the overhead associated with training full-scale models. Despite their rapid proliferation and demonstrated success in natural language processing \cite{han2024parameter}, a comprehensive understanding of how PEFT methods are applied in software engineering remains lacking. Key questions persist: \textit{Which PEFT strategies are favored for adapting LCMs in software engineering? -- Are certain model architectures or task types better suited for PEFT? What trade-offs exist? \etc}
In short, the practical effectiveness and limitations of PEFT techniques in the context of code intelligence have yet to be thoroughly examined.

In this paper, we present a Systematic Literature Review (SLR) characterizing generative and non-generative software engineering-related tasks.
The research questions we aim to answer, will highlight three different dimensions concerning PEFT-optimization methods for SE-related tasks, these include: RQ$_{1}$ the types of software engineering tasks that have been targeted using PEFT techniques; RQ$_{2}$ the types of LLM architecture (\eg encoder-decoder) and PEFT methods commonly employed in software engineering applications, and finally; RQ$_{3}$ the impact on task performance and training efficiency of PEFT-based techniques in generative and non-generative SE tasks, when compared against valid full fine-tuning baselines.

By systematically answering these questions, our study aims to bridge the gap in understanding PEFT’s role in optimizing LCMs for software engineering. We provide a comprehensive taxonomy of PEFT applications in SE and distill empirical insights that can guide both future research and practical adoption of these techniques in AI-driven software development environments. This is accomplished through a comprehensive analysis of the current landscape of efficient model adaptation strategies for both generative and non-generative SE tasks, highlighting how PEFT techniques can make the training and deployment of Large Code Models more practical and sustainable in real-world development settings.

The remainder of this paper is organized as follows: \secref{subsec:lcm} introduces the use of LLMs in software engineering\footnote{Note that in this investigation, the terms LLM4SE (Large Language Models for Software Engineering) and LCMs (Large Code Models) are used interchangeably, as both refer to deep learning models tailored for automating software engineering tasks through code understanding and generation.}, providing an overview of current practices in SE automation and detailing recent advancements at the overlap of Artificial Intelligence (AI) and SE, a field commonly known with the word AI4SE. This section also examines the evolution and variants of LCMs, with a particular focus on their architectural characteristics, which—as we will show—play a key role in determining the choice of suitable PEFT optimization strategies. These foundational elements are further discussed in \secref{subsec:peft}, which delves into the principles behind PEFT techniques.

We then present the methodology of our study in \secref{sec:methodology}, outlining our approach to data collection, paper selection, and classification. \secref{sec:results} provides detailed answers to each of our three research questions, offering insights drawn from the systematic review. In \secref{sec:future}, we discuss the implications of our findings and suggest promising directions for future research on the application of PEFT in SE.

Finally, we address potential threats to the validity of our study in \secref{sec:threats}, and conclude with a summary of our key takeaways and reflections in \secref{sec:conclusion}. Central to the contributions of this SLR is the taxonomy introduced in \figref{fig:peft_se_taxonomy}, which synthesizes our findings into a structured classification of PEFT applications across generative and non-generative software engineering tasks. This taxonomy not only distills patterns observed throughout the review but also serves as a practical framework for guiding future applications of PEFT. By clarifying the alignment between PEFT strategies and software engineering task types, the taxonomy equips researchers and practitioners with a clearer understanding of the current landscape—ultimately enabling a more confident and precise application as well as design of future parameter-efficient adaptation strategies.


\section{Background and Related Work}
\label{sec:background}
In this section, we first provide an overview of recent advancements in LCMs and their impact on various software engineering tasks. We then highlight the evolution of PEFT techniques, detailing their methodologies, advantages, and applications in optimizing LLMs.

\subsection{Large Code Models}
\label{subsec:lcm}

Recent advancements in Large Code Models (LCMs) have significantly accelerated the automation of software engineering tasks, contributing to notable improvements in both efficiency and accuracy across development pipelines \cite{hou2024large, niu2022deep, niu2023empirical}. The remarkable capabilities of LCMs are rooted in their large-scale pre-training on extensive and diverse textual and code corpora, as well as the substantial parameter counts that enhance their learning capacity.

While general-purpose LLMs, such as GPT-4 \cite{achiam2023gpt} and Claude \cite{Anthropic_Claude3}, exhibit broad applicability across multiple domains, their generality often limits their effectiveness in code-intensive tasks that require domain-specific knowledge. To address this challenge, researchers have introduced specialized LCMs that are explicitly designed to meet the unique demands of software engineering applications. As a result, LCMs have been successfully deployed across a diverse range of SE tasks, including but not limited to Code Generation \cite{fan2023large, li2022competition, codegen}, Code Summarization \cite{gu2022assemble, wang2023one}, Code Clone Detection \cite{svajlenko2016bigcloneeval, white:ase2016}, Bug Fixing \cite{tufano2020unit, chen2023teaching}, Requirements Analysis \cite{marques2024using} and Automated Program Repair \cite{jiang:icse2021, mashhadi:msr2021}. These models build upon the Transformer architecture introduced by Vaswani \etal ~\cite{vaswani:nips2017} and are typically categorized under three primary architectural frameworks: encoder-only, encoder-decoder, and decoder-only models.

\textbf{Encoder-only} models, such as BERT \cite{devlin2018bert}, are based on the Transformer encoder and are frequently adapted for software engineering use cases. Variants such as CodeBERT \cite{feng2020codebert}, GraphCodeBERT \cite{guo:iclr2021}, RoBERTa \cite{liu:arxiv2019}, and ALBERT \cite{lan2019albert} have been widely employed in SE tasks. Additionally, specialized models such as BERTOverflow \cite{tabassum2020code} and CodeRetriever \cite{li2022coderetriever} have been designed to capture the structural and semantic nuances of source code by introducing tailored pre-training objectives or by integrating program structure representations. For example, CodeBERT employs token prediction to enhance programming language understanding for downstream tasks such as Code Completion and Documentation Generation \cite{feng2020codebert}. Furthermore, Salza \etal \cite{salza2022effectiveness} evaluated the capacity of encoder-based models (\ie BERT, RoBERTa) to capture both natural language and source code semantics, ultimately improving Code Search and Retrieval tasks.

\textbf{Encoder-decoder} models, including PLBART \cite{ahmad2021unified}, T5 \cite{raffel2020exploring}, and CodeT5 \cite{wang2021codet5identifierawareunifiedpretrained}, leverage bidirectional encoding and autoregressive decoding to address a variety of SE challenges. Extensions such as CodeT5+ \cite{wang2023codet5+}, AlphaCode \cite{li2022competition}, and CoTexT \cite{phan2021cotext} have demonstrated enhanced adaptability in tasks such as Bug Fixing and Code Summarization \cite{al2023extending, gu2022assemble, mastropaolo:icse2021}. Additionally, ATHENATEST, proposed by Tufano \etal \cite{tufano2020unit}, adapts the BART framework \cite{lewis2019bart} to generate unit tests, showcasing the versatility of encoder-decoder architectures in supporting software quality assurance tasks.

\textbf{Decoder-only} models, such as PaLM \cite{chowdhery2023palm}, GPT-3.5 \cite{openai_gpt35}, GPT-4 \cite{achiam2023gpt}, and LLaMA \cite{touvron2023llama}, represent a further evolution in LLM design, offering highly effective generative capabilities. For instance, Codex, with 12 billion parameters, has demonstrated the ability to solve 72.31\% of complex Python programming problems and serves as the backbone for tools such as GitHub Copilot \cite{chen2021evaluating, minaee2024large}. GPT-4 continues to set benchmarks across multiple SE tasks, including Code Synthesis, Debugging, and Reasoning \cite{bubeck2023sparks}. Evaluations by Arakelyan \etal \cite{arakelyan2023exploring} on Codex and CodeT5 have further validated their effectiveness in Code Generation and Code Summarization, even under distribution shifts.
Among the specialized decoder-only models, CodeLlama is a prominent family of open-source LLMs derived from LLaMA 2 \cite{touvron2023llama}. It has been further trained on a corpus of 500 billion tokens, including both code and natural language, making it highly effective for various code-related tasks \cite{zan2024codes, chen2021evaluating, xia2023universal, virk2024enhancing}. Likewise, DeepSeek-Coder \cite{deepseekai2024deepseekcoderv2breakingbarrierclosedsource} has emerged as a strong competitor, featuring models ranging from 1B to 33B parameters that have been pre-trained on 2 trillion tokens, including extensive code-related data. Remarkably, DeepSeek-Coder has been shown to outperform larger models like GPT-3.5 \cite{brown2020language}, with its 6.7B parameter model even surpassing the 33B variant of CodeLlama in certain benchmarks.

The effectiveness of these models is further augmented by continuous improvements in both architecture and task-specific tuning methods. Innovations such as infilling objectives \cite{li2023starcoder} and instruction tuning \cite{xu2024wizardlm} have been integrated into models like StarCoder and CodeLlama, improving their performance on tasks such as Code Completion and Code Editing \cite{roziere2023code}. Additionally, LCMs have been effectively applied to domain-specific and low-resource programming languages, including Verilog \cite{liu2023verilogevalevaluatinglargelanguage} and Haskell \cite{vandam2024investigatingperformancelanguagemodels}, often utilizing transfer learning from high-resource languages like Python and Java to overcome data scarcity issues \cite{zhang2023doesllmgeneratesecurity}.
Beyond their conventional applications in programming tasks, LCMs have also been leveraged in emerging areas such as Autonomous Agent Systems for workflow automation \cite{hong2024metagptmetaprogrammingmultiagent} and in assessing the quality of AI-generated code \cite{wei2023towards}.

In conclusion, LCMs have consistently demonstrated state-of-the-art performance across a wide range of SE tasks, including Requirements Analysis \cite{marques2024using}, Code Comprehension \cite{10.1145/3597503.3639138, ding2023staticevaluationcodecompletion}, Code Generation \cite{li2022competition, mastropaolo2024log}, Code Translation \cite{pan2023stelocoder}, Code Summarization \cite{sun2024source, mastropaolo2024towards, mastropaolo2022using}, Automated Program Repair \cite{fan2023automated, jiang2023impactcodelanguagemodels, 10.1145/3618305.3623587, xia2023automated}, Code Smell Detection and various Software Testing activities \cite{article,tufano2020unit,article}. The continued evolution of models such as CodeLlama, DeepSeek-Coder, and CodeT5 underscores the transformative impact of domain-specific LCMs in automating and enhancing modern software engineering practices.

\subsection{Fine-Tuning Code Models}
\label{subsec:FT}
The effectiveness of LCMs largely depends on their architectural scale, which often includes millions to billions of parameters. These models undergo pretraining on extensive source code corpora before being further adapted to downstream tasks through fine-tuning. Fine-tuning strategies can be broadly categorized as follows:

\textbf{Full Fine-Tuning (FFT):}
This method initializes a Pretrained Language Model (PLM) with pretrained weights and refines it using task-specific data via backpropagation and gradient descent \cite{tayaranian2022towards}. In FFT, all model parameters, including the pretrained weights, are updated to minimize a task-specific loss function that quantifies the deviation between model predictions and ground truth. However, full fine-tuning is computationally intensive, requiring substantial computational resources and large labeled datasets. The challenge becomes even more pronounced as PLMs scale to billions of parameters, making FFT impractical for many applications \cite{patil2024review}.

\textbf{Parameter-Efficient Fine-Tuning (PEFT):}
To address the resource-intensive nature of FFT, researchers have developed PEFT methods, which update only a small subset of model parameters while keeping the rest frozen \cite{houlsby2019parameterefficienttransferlearningnlp}. This substantially reduces both memory requirements and computational costs, making it a more feasible approach for adapting LLMs to specific tasks. Recent studies have demonstrated the effectiveness of PEFT in software engineering applications, including Code Generation \cite{weyssow2023exploring}, Code Review \cite{lu2023llama}, Code Clone Detection \cite{ayupov2022parameter}, and Automated Program Repair \cite{li2024exploring}. These studies highlight how PEFT significantly reduces training time and memory consumption while maintaining or even improving model performance in various software engineering tasks.


\subsection{Parameter-Efficient Fine-Tuning (PEFT)}
\label{subsec:peft}
PEFT techniques can be categorized into \textit{Additive}, \textit{Selective}, \textit{Reparameterized}, and \textit{Hybrid} fine-tuning based on their underlying mechanisms. 

\subsubsection{ \textbf{Additive PEFT:}} Additive PEFT methods introduce new trainable parameters while keeping the original model parameters frozen, making them the most widely explored category in PEFT research. Two major subcategories within this approach include Adapter-based methods and Soft Prompts.

\textbf{Adapter-based} approaches integrate small adapter layers within Transformer blocks to enable task-specific adaptation without modifying the core model parameters \cite{houlsby2019parameterefficienttransferlearningnlp}.
An adapter layer typically comprises a down-projection matrix $\mathbf{W}_{\text{down}} \in \mathbb{R}^{r \times d}$, followed by a non-linear activation function $\sigma(\cdot)$, and an up-projection matrix $\mathbf{W}_{\text{up}} \in \mathbb{R}^{d \times r}$. Here, $d$ denotes the dimensionality of the hidden layer, while $r$ is the bottleneck dimension, a tunable hyperparameter that controls the adapter’s capacity. Given an input $h_{\text{in}}$ to the adapter, the transformation performed by the adapter layer (including the residual connection) can be expressed as:
\[
\text{Adapter}(x) = \mathbf{W}_{\text{up}} \sigma(\mathbf{W}_{\text{down}} x) + x.
\]
This methodology has been widely adopted \cite{houlsby2019parameterefficienttransferlearningnlp, ruckle2020adapterdrop, lin2020exploring, lei2023conditional, he2021towards} and has led to multiple variations, such as modifications in adapter placement for instance SparseAdapter \cite{he2022sparseadapter}, Pruning techniques to enhance efficiency \cite{he2021towards}, and Reparameterization strategies aimed at reducing the number of trainable parameters \cite{karimi2021compacter}.

Two prominent types of adapters are the \textbf{Language Adapter (L-Adapter)} and the \textbf{Task Adapter (T-Adapter)} \cite{goel2022crossmodaltransfernaturallanguage}. The L-Adapter is trained using a masked language modeling objective on unlabeled data from a specific target language, thereby enabling the PLMs to learn language-specific characteristics. In contrast, the T-Adapter is optimized on labeled data to perform a specific downstream task, facilitating task-specific adaptation. This modular design enables PLMs to be effectively adapted to previously unseen languages, even those not originally covered during the model's pretraining phase.

\textbf{Named Entity Recognition (NER)} \cite{saberi2024utilizationpretrainedlanguagemodel} Adapters are lightweight modules integrated within the Transformer architecture. Their primary objective is to capture and encode \textit{type information} derived from the Abstract Syntax Tree (AST) of source code. By incorporating structural and semantic cues from the AST, these adapters enhance the model’s ability to recognize and classify named entities within code, contributing to more accurate and context-aware representations.

\textbf{AdapterFusion}~\cite{pfeiffer2020adapterfusion} is a framework that integrates multiple task-specific adapters into a unified adapter module, facilitating effective knowledge transfer across related tasks without altering the parameters of the original pretrained model. This approach enables efficient task composition, allowing pretrained language models to be reused across a variety of downstream tasks.

\textbf{MAM Adapter}~\cite{he2021towards} explores the intrinsic similarities among three additive Parameter-Efficient Fine-Tuning (PEFT) methods: adapters, prefix-tuning, and LoRA. This insight leads to the development of three adapter variants. 
The first variant, \textit{Parallel Adapter}, places adapter layers alongside specific components of the model, such as the self-attention (SA) or feedforward network (FFN) layers, instead of inserting them sequentially after these components. The second, \textit{Multi-head Parallel Adapter}, divides the parallel adapter into multiple heads, with each head independently modifying the output of a corresponding attention head in the SA module. The third, \textit{Scaled Parallel Adapter}, introduces a scaling factor following the parallel adapter layer, aligning with the design principles of LoRA.

\textbf{Soft Prompt} Tuning, another prominent additive PEFT method, introduces trainable continuous vectors, referred to as soft prompts, into either the input representation or the hidden states of the model.
This can be represented as follows:
\[
X_{plq} = \left( s_{plq_1}, \dots, s_{plq_{N_S}}, x_{plq_1}, \dots, x_{plq_{N_X}} \right)
\]
where \(X_{plq}\) is the sequence of input tokens for layer \(l\), including soft prompt tokens \(s_{plq_i}\) followed by the original input tokens \(x_{plq_i}\). \(N_S\) is the number of soft prompt tokens, and \(N_X\) is the number of original input tokens.
Unlike manually crafted hard prompts—static natural language instructions or templates written by humans—soft prompts are automatically optimized through task-specific training data, allowing the model to learn an optimal prompt configuration. Variants of this approach include \textbf{Prefix Tuning}, which pre-pends soft prompts to the hidden states of the multi-head attention mechanism \cite{li2021prefix}, and \textbf{Prompt Tuning}, which appends learnable prompt tokens to the input embeddings while ensuring that only these prompt-related parameters are updated during training \cite{lester2021power}. By keeping the core model parameters frozen, these techniques offer computationally efficient fine-tuning while maintaining strong task adaptability.

\textbf{P-Tuning}~\cite{liu2024gpt} introduces a prompt-based fine-tuning strategy by inserting a sequence of continuous soft prompts, denoted as $[P_1], \cdots, [P_i], [P_{i+1}], \cdots, [P_l]$, into the input of a pretrained language model. Unlike traditional Prompt Tuning, P-Tuning constructs a learnable template by concatenating these prompts and mapping them into the embedding space as follows:  
\[
\{h_1, \cdots, h_i, e(x), h_{i+1}, \cdots, h_l, e(x)\},
\]
where $e(x)$ represents the output from the pretrained embedding layer and $\{h_1, \cdots, h_l\}$ are the trainable continuous prompt representations. 
The training objective is to optimize these soft prompt embeddings while keeping the parameters of the PLM fixed. This approach significantly reduces the number of trainable parameters, making it particularly effective in few-shot learning scenarios.

\textbf{Pass-Tuning} \cite{chen-etal-2023-pass} is a structure-aware, parameter-efficient fine-tuning strategy specifically designed for code representation learning. This approach introduces a modular, plug-and-play Graph Neural Network (GNN) component, which functions as a tunable prefix and is trained to capture structural information from the Abstract Syntax Tree of source code. By explicitly modeling the syntactic and hierarchical relationships inherent in the AST, Pass-Tuning facilitates a more nuanced and semantically rich understanding of code structure. This enables the underlying language model to better encode structural dependencies, thereby enhancing the quality of the learned code representations while maintaining efficiency in parameter usage.

\textbf{(IA)$^3$ (Infused Adapter by Inhibiting and Amplifying Inner Activations)}~\cite{liu2022few} introduces an efficient parameter-efficient tuning method that enhances model performance by modifying internal activations through learned scaling vectors. Specifically, (IA)$^3$ injects three learned vectors—$l_k$, $l_v$, and $l_{\text{ff}}$—into the model. These vectors are used to rescale the key ($K$) and value ($V$) projections in the self-attention mechanism, as well as the hidden activations within the position-wise feedforward network (FFN).

AdaMix \cite{wang2022adamix}, AdapterDrop \cite{ruckle2020adapterdrop}, Ladder-Side Tuning (LST) \cite{sung2022lst}, and Intrinsic Prompt Tuning (IPT) \cite{qin2021exploring} are other popular additive PEFT methods.

\subsubsection{\textbf{Selective PEFT: }}
Selective PEFT optimizes a subset of existing parameters rather than introducing additional components, reducing computational overhead while maintaining adaptability. 
Consider a model with parameters $\theta = \{\theta_1, \theta_2, \ldots, \theta_n\}$, where each $\theta_i$ denotes an individual model parameter, and $n$ is the total number of parameters. Selective PEFT operates by applying a binary mask $M = \{m_1, m_2, \ldots, m_n\}$ over these parameters, where each $m_i \in \{0, 1\}$ indicates whether parameter $\theta_i$ is selected ($m_i = 1$) or excluded ($m_i = 0$) from fine-tuning. The updated parameter set $\theta'$ after fine-tuning can be expressed as:
\[
\theta_i' = \theta_i - \eta \cdot m_i \cdot \frac{\partial L}{\partial \theta_i},
\]
where $\eta$ is the learning rate, and $\frac{\partial L}{\partial \theta_i}$ represents the gradient of the loss function $L$ with respect to $\theta_i$. Under this formulation, only the parameters for which $m_i = 1$ are updated during backpropagation, while the remaining parameters remain fixed.

It can be classified into three approaches: \textit{Bias update} methods, such as \textbf{BitFit (Bias-term Fine-tuning)} \cite{zaken2021bitfit}, which fine-tunes only bias terms and task-specific layers while keeping most parameters frozen; \textit{pretrained weight masking}, such as \textbf{FISH Mask (Fisher-Induced Sparse Hanging)} \cite{sung2021training}, which applies structured pruning to mask pretrained weights; and \textit{delta weight masking} \cite{ansell2021composable, xia2022towards, guo2020parameter}, which prunes delta weights using optimization techniques like \textbf{LT-SFT (Lottery Ticket Sparse Fine-Tuning)} \cite{ansell2021composable}. 

For Instance, \textbf{Telly-K} \cite{shi2023towards} method uses a layer-freezing strategy where the bottom $K$ layers of the model are frozen— \ie their parameters remain unchanged—while only the top $(L - K)$ layers are updated during training, where $L$ denotes the total number of transformer layers. For example, in the \textit{Telly-6} configuration, layers $0$ through $5$ are fixed, and fine-tuning is applied exclusively to layers $6$ through $11$.

\subsubsection{\textbf{Reparameterized PEFT:}}
Reparameterized PEFT methods employ low-rank transformations to reduce the number of trainable parameters while preserving the model's ability to operate on high-dimensional matrices, such as pretrained weights. This category includes \textit{intrinsic dimension-based} approaches, such as \textbf{Intrinsic SAID} \cite{aghajanyan2020intrinsic}, which was one of the earliest techniques exploring the intrinsic dimensionality of fine-tuning LLMs. However, the most widely adopted reparameterization technique is \textbf{LoRA (Low-Rank Adaptation)} \cite{hu2022lora}. LoRA optimizes fine-tuning by introducing two trainable low-rank matrices—a down-projection matrix and an up-projection matrix—which are applied in parallel to update model weights efficiently. 
Given a pre-trained weight matrix $\mathbf{W}_0 \in \mathbb{R}^{d \times k}$, LoRA (Low-Rank Adaptation) introduces two additional trainable matrices: a down-projection matrix $\mathbf{W}_{\text{down}} \in \mathbb{R}^{r \times k}$ and an up-projection matrix $\mathbf{W}_{\text{up}} \in \mathbb{R}^{d \times r}$, where the rank $r \ll \min(d, k)$. These matrices operate in parallel to $\mathbf{W}_0$, forming a low-rank approximation of the weight update. Letting $h_{\text{in}}$ represent the input, the standard output is computed as \[
h_{\text{out}} = \mathbf{W}_0 h_{\text{in}}.
\]
LoRA modifies this computation by injecting a task-specific incremental update $\Delta \mathbf{W}$, resulting in:
\[
h_{\text{out}} = \mathbf{W}_0 h_{\text{in}} + \alpha \mathbf{W}_{\text{up}} \mathbf{W}_{\text{down}} h_{\text{in}},
\]
where $\alpha$ is a scaling factor that controls the magnitude of the adaptation.
During initialization, $\mathbf{W}_{\text{down}}$ is sampled from a Gaussian distribution, while $\mathbf{W}_{\text{up}}$ is initialized to zero, ensuring that $\Delta \mathbf{W} = 0$ at the start of training. This design makes LoRA straightforward to implement, with demonstrated scalability on models with up to 175 billion parameters. Once fine-tuning is completed, the learned LoRA weights can be merged with the pre-trained weights $\mathbf{W}_0$, allowing the model to retain its efficiency and avoid additional inference overhead.

Several LoRA-based extensions have been proposed, including \textbf{KronA (Kronecker Adapter)} \cite{edalati2022krona}, \textbf{DyLoRA} \cite{valipour2022dylora} , \textbf{GLoRA} \cite{chavan2023one}, and \textbf{AdaLoRA} \cite{zhang2023adalora}, each introducing variations to enhance efficiency and task adaptability.

\textbf{AdaLoRA (Adaptive Low-Rank Adaptation)}~\cite{zhang2023adalora} extends the standard LoRA framework by introducing a mechanism for dynamically adjusting the rank of adaptation matrices, thereby enabling efficient control over parameter allocation budgets. In this approach, the incremental weight update $\Delta W$ is reparameterized using Singular Value Decomposition (SVD) as $\Delta W = P \Lambda Q$, where $P$ and $Q$ are orthogonal matrices, and $\Lambda$ is a diagonal matrix containing the singular values $\{\sigma_1, \sigma_2, \ldots, \sigma_r\}$. The variable $r$ denotes the rank of $\Lambda$.
During training, the matrices $P$ and $Q$ are initialized with values sampled from a Gaussian distribution and regularized to preserve orthogonality, while $\Lambda$ is initialized to zero. The rank of $\Lambda$ is iteratively pruned by truncating its smaller singular values, allowing the model to adaptively control the complexity of $\Delta W$.

\textbf{FF-LoRA} \cite{he2021towards} is a variant of the LoRA framework that applies the low-rank update mechanism specifically to the \textit{Feed-Forward} (FF) layers of Transformer-based architectures. While the original LoRA approach introduces a learnable low-rank decomposition $\Delta W = AB$, added to the model weights post-finetuning, its application was initially restricted to the \textit{Query} and \textit{Value} projections within the Multi-Head Attention mechanism. FF-LoRA extends this methodology by applying the low-rank update to the FF layers, enabling parameter-efficient adaptation in a different architectural component.

A notable extension of LoRA, \textbf{QLoRA} (Quantized LoRA) \cite{dettmers2023qlora}, integrates quantization techniques to further reduce computational demands. QLoRA fine-tunes transformer models by applying 4-bit NormalFloat (NF4) quantization and double quantization processing, significantly improving memory efficiency while maintaining fine-tuning effectiveness.

\subsubsection{\textbf{Hybrid/Others Technique:}}
Hybrid fine-tuning combines multiple PEFT techniques, such as Adapters, Prefix Tuning, and LoRA, to maximize efficiency and adaptability. Notable approaches include \textbf{MAM Adapter} \cite{he2021towards}, \textbf{Compacter} \cite{karimi2021compacter}, and \textbf{AutoPEFT} \cite{zhou2024autopeft}, which integrate diverse strategies for enhanced performance.

For Instance, \textbf{MAM Adapter (Mix-And-Match)}~\cite{he2021towards} explores the intrinsic similarities among three PEFT methods: adapters, Prefix Tuning, and LoRA. This insight leads to the development of three adapter variants. 
The first variant, \textit{Parallel Adapter}, places adapter layers alongside specific components of the model, such as the self-attention (SA) or feedforward network (FFN) layers, instead of inserting them sequentially after these components. The second, \textit{Multi-head Parallel Adapter}, divides the parallel adapter into multiple heads, with each head independently modifying the output of a corresponding attention head in the SA module. The third, \textit{Scaled Parallel Adapter}, introduces a scaling factor following the parallel adapter layer, aligning with the design principles of LoRA.

Prior research has explored the application of PEFT techniques across various NLP tasks. Han \etal \cite{han2024parameter} conducted a comprehensive literature survey on existing PEFT methods and their adaptation across different domains. Lialin \etal \cite{lialin2023scaling} performed a comprehensive experimental evaluation of PEFT methods, benchmarking 14 techniques and their variants across five datasets and three model scales (0.7B, 3B, and 11B parameters). The study provides an in-depth analysis of the methods’ efficiency, focusing on GPU memory usage and throughput performance. Other studies \cite{lialin2023scaling, xu2023parameter} have similarly examined PEFT frameworks from a broader perspective. However, a systematic review of PEFT techniques specifically within software engineering remains largely unexplored. This study aims to systematically collect, analyze, and synthesize research on PEFT applications in SE tasks, providing a structured understanding of their effectiveness and impact. Ultimately, this work provides actionable insights for both researchers and practitioners by identifying which PEFT strategies are most effective across various SE tasks, outlining current limitations, and highlighting promising avenues for future research on integrating efficient fine-tuning techniques into software engineering workflows.


\section{Methodology: Study Selection and Analysis}
\label{sec:methodology}

In this section, we outline the structure of our systematic literature review, following the well-established guidelines proposed by Kitchenham \etal \cite{kitchenham2009systematic}.

Our review is driven by three key Research Questions (RQs), which shape the direction of our investigation and ensure a focused analysis of primary studies related to PEFT techniques in software engineering.
Specifically, our SLR is guided by the following RQs:

\begin{itemize}
    \item \textbf{RQ\textsubscript{1}: \textit{Which software engineering tasks have been explored using PEFT methods?}}  \\
    This research question investigates the full spectrum of software engineering applications where PEFT techniques have been applied, providing an overview of their adoption in the field. In particular, our focus is on highlighting differences (if any) that may emerge across various types of SE tasks, such as generative tasks (\eg code generation, summarization) versus non-generative tasks (\eg code classification, clone detection). By examining how PEFT strategies are utilized in these distinct contexts, we aim to uncover patterns of usage, task-specific adaptations, and potential gaps in the current literature that warrant further exploration.\\

    \item \textbf{RQ\textsubscript{2}: \textit{Which deep learning models and architectures are commonly optimized with PEFT methods?}}  \\
    This research question aims to identify the types of base models and architectural families that are commonly fine-tuned using PEFT techniques. Additionally, it explores how different PEFT methods are instantiated across various model types (\eg encoder-only, decoder-only, encoder-decoder) to support software engineering tasks. The goal is to understand which architectural configurations are most ``compatible'' with PEFT optimization methods.\\

    \item \textbf{RQ\textsubscript{3}: \textit{How do PEFT methods impact task performance and training efficiency in software engineering tasks when compared against full fine-tuning?}} \\
    This research question investigates the impact of the best-performing PEFT method(s) on both task performance (\eg BLEU, Accuracy) and efficiency (\eg training time, memory usage) across generative and non-generative software engineering tasks. To ensure a fair comparison, we consider only those studies that present a one-to-one evaluation between PEFT-optimized models and their fully fine-tuned counterparts. This approach enables us to draw meaningful insights into the practical applicability of PEFT techniques in real-world SE settings. For each study, we record whether the selected PEFT method leads to improvements, degradations, or neutral effects in terms of both performance and efficiency.

\end{itemize}

Through our examination of these research questions, we aim to map the landscape of Parameter-Efficient Fine-Tuning in software engineering, highlighting both its capabilities and limitations. 
Our review methodology consists of the following steps:

\begin{enumerate}
    \item \textbf{Identifying and Selecting Primary Studies}
    \begin{enumerate}
        \item Executing search queries across four databases using a defined publication year range (2019–2025).
        \item Manually screening results for relevance to PEFT methods and SE tasks.
    \end{enumerate}

    \item \textbf{Study Selection and Filtering Process}
    \begin{enumerate}
        \item \textit{Exclusion Criteria and Alignment Analysis:} Applied exclusion criteria and conducted manual review to finalize 25 studies, resolving disagreements through consensus.
        \item \textit{Snowballing and Manual Addition of Studies:} Identified and included two additional studies through backward snowballing, resulting in a final set of 28 papers.
    \end{enumerate}
    \item \textbf{Data Extraction and Synthesis}
    \begin{enumerate}
        \item \textit{Data Extraction:} Conducted a high-level extraction of relevant study attributes aligned with our research questions.
        \item \textit{Synthesis Approach:} Categorized SE tasks as generative or non-generative and analyzed trends in PEFT application across task types and deep learning models, as well as architectures.
    \end{enumerate}
\end{enumerate}

\begin{figure}[h]
\centering
\includegraphics[width=\textwidth]{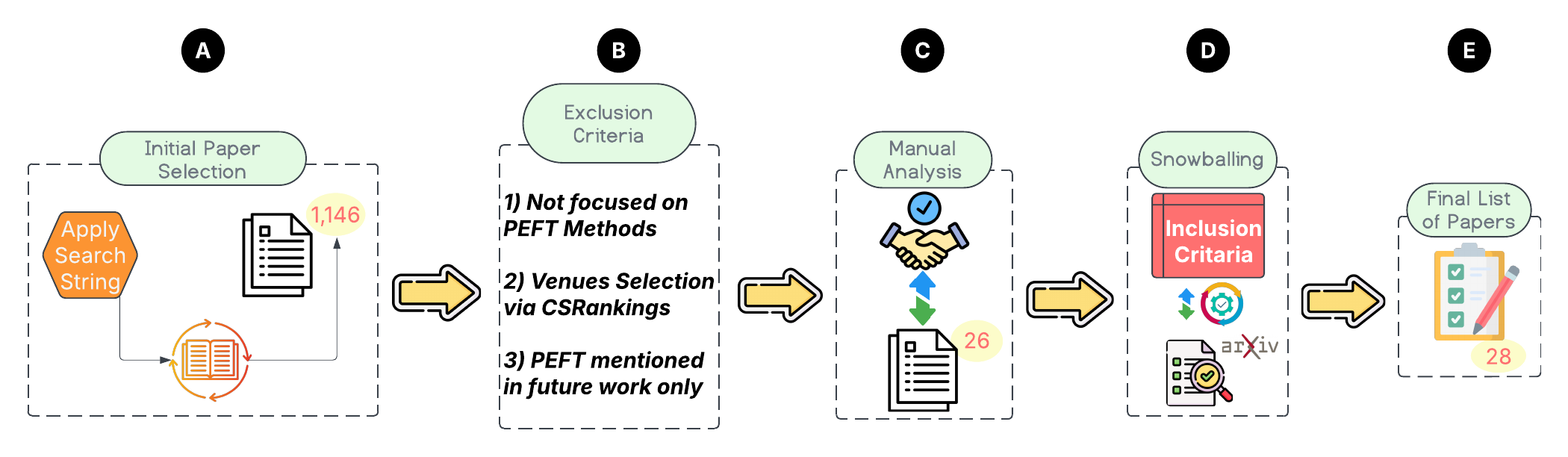}
\caption{Paper Selection Process for the Systematic Literature Review on PEFT Techniques in Software Engineering Tasks. The pipeline consists of five stages: applying search strings to databases—\protect\circledletter{A}, exclusion criteria—\protect\circledletter{B}, manual analysis—\protect\circledletter{C}, snowballing—\protect\circledletter{D}, and final selection—\protect\circledletter{E}.}
\label{fig:slr_pipeline} 
\end{figure}

\figref{fig:slr_pipeline} illustrates our paper selection workflow, outlining the progression from the initial search to the final set of papers included in our SLR. The detailed methodology for each step in this process is described in \secref{subsec:filtering_process} (Study Selection and Filtering Process), including our search strategy, venue selection, and the application of inclusion and exclusion criteria such as snowballing.

\subsection{Identifying and Selecting Primary Studies} \label{subsec:study_selection}

To properly address the RQs underpinning our investigation, we look for primary studies by defining a time frame, ranging from 2019 to 2025, which aligns with the rise and evolution of PEFT techniques as a widely adopted method for optimizing large models.


Next, we began our search by querying four major electronic databases \textbf{IEEE Xplore}, \textbf{ACM Digital Library}, \textbf{Springer Link}, and \textbf{Google Scholar} (via \textit{Publish or Perish} software) using a tailored search string designed to capture studies involving PEFT in SE contexts. This initial step--depicted in \figref{fig:slr_pipeline}-\circled{A}, resulted in \textbf{1,146} forming the basis of our analysis\footnote{All the retrieved studies are available for consultation in our online appendix \cite{replication} studies forming the basis of our analysis.}.

After retrieving the initial set of studies, we applied a two-stage filtering process.
First, we conducted a relevance screening to ensure that the papers explicitly addressed PEFT methods and their application to software engineering tasks. This screening was performed by the first author, who manually reviewed the initial pool of 1,146 studies. In cases of uncertainty regarding the relevance of a particular approach to the software engineering domain, the first author consulted with the second author. If disagreement persisted, the conflict was solved by including the third author of the investigation. This allowed us to reach a consensus. 
Next, we applied \textit{venue-based filtering} to ensure the quality and impact of the selected research. To this end, we focused on top-tier conferences and journals across software engineering, programming languages, machine learning, and natural language processing--\figref{fig:slr_pipeline}-\circled{B}.
Our venue selection was informed by \textbf{CSRankings}\footnote{\url{https://csrankings.org}}, a metrics-driven and widely accepted ranking of top computer science venues. From this source, we selected high-impact venues including: \\

\begin{itemize}
    \item \textbf{Artificial Intelligence}: AAAI, IJCAI
    \item \textbf{Machine Learning}: ICLR, ICML, NeurIPS, KDD
    \item \textbf{Natural Language Processing}: ACL, EMNLP, NAACL
    \item \textbf{Programming Languages}: PLDI, POPL
    \item \textbf{Software Engineering}: FSE, ICSE, ASE, ISSTA
\end{itemize}

In addition to the venues extracted from CSRanking, we also included several well-regarded software engineering conferences and journals that are not listed in CSRankings but are recognized in the community for their relevance and peer-reviewed contributions. These include journals such as the ACM Transactions on Software Engineering and Methodology (TOSEM), IEEE Transactions on Software Engineering (TSE), and Empirical Software Engineering (EMSE), and conferences such as the Mining Software Repositories (MSR), International Conference on Software Analysis, Evolution and Reengineering (SANER), International Conference on Program Comprehension (ICPC), the international conference on AI Foundation Models and Software Engineering (FORGE).

\begin{figure*}[h]
\centering
\includegraphics[width=0.8\textwidth]{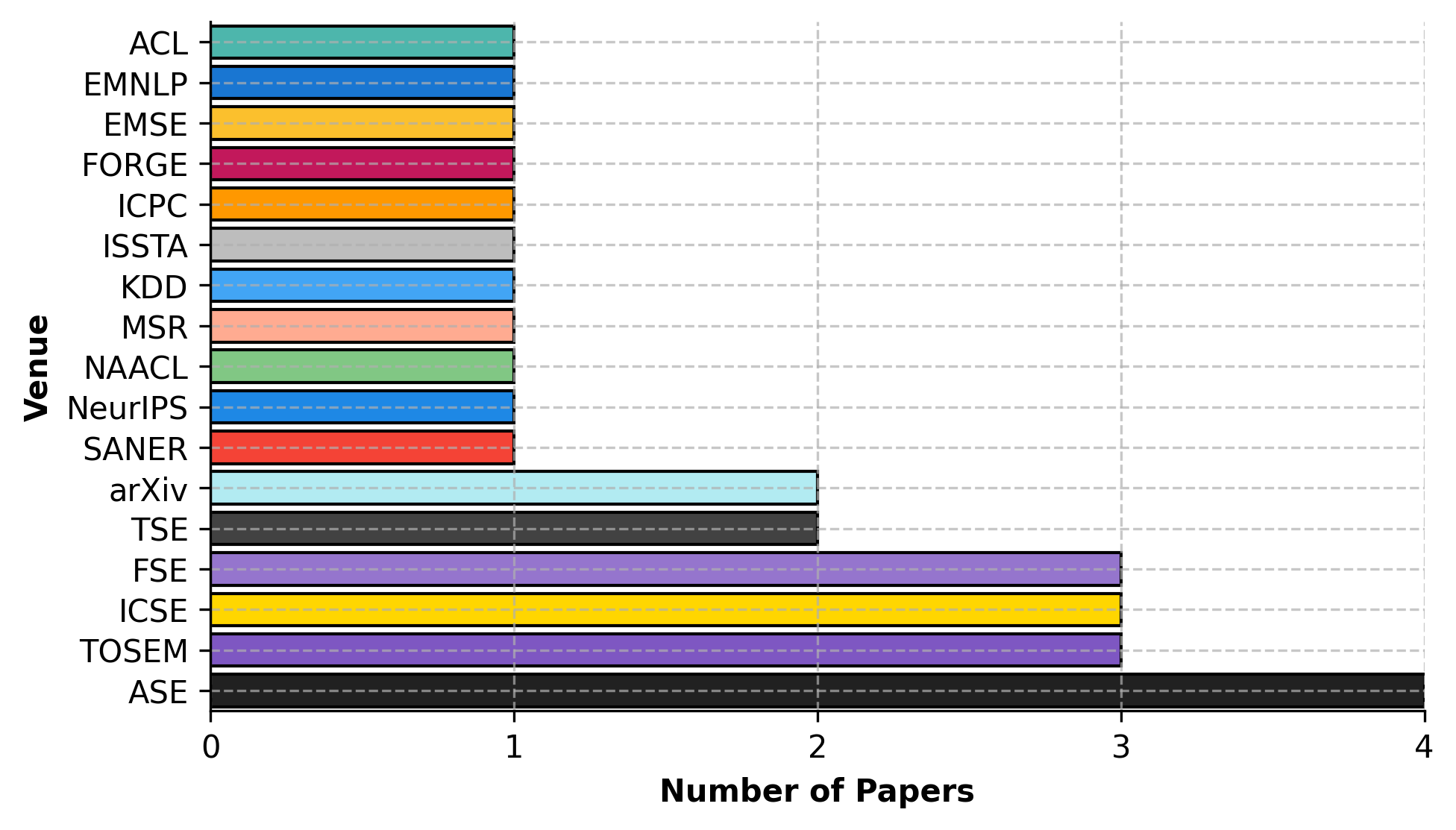}
\caption{Venue Distribution of PEFT Techniques in Software Engineering.}
\label{fig:venue_selection}
\end{figure*}

As shown in \figref{fig:venue_selection}, our final collection of papers spans 16 venues, with the ASE, ICSE, and TOSEM contributing the largest number of relevant studies. While this approach may exclude some potentially relevant papers from less prominent venues, it ensures our review emphasizes high-quality, peer-reviewed research from both top-ranked and widely recognized publication forums.

To further ensure rigor, we deliberately excluded preprint repositories (\eg \textit{arXiv}) from our primary database search. Although such sources may contain cutting-edge findings, the lack of formal peer review could introduce validity threats. Nonetheless, let us anticipate that through our \textit{snowballing process}, we considered making an exception for two highly cited arXiv papers \cite{ayupov2022parameter, goel2022crossmodaltransfernaturallanguage} that had substantial influence in the peer-reviewed literature, demonstrating their relevance and impact.
In doing so, we adopted an approach grounded in sound principles and informed by domain expertise—allowing us to holistically assess the quality of each study. This includes evaluating its scholarly impact, such as citation count, and its influence within the relevant research community.

With a general overview of the approach we followed to implement the SLR, let us now introduce the fundamental details underlying the retrieval phase--\figref{fig:slr_pipeline}--\circled{A}.

We first design the search string based on key terms extracted from our RQs, ensuring comprehensive coverage of PEFT methods (\eg LoRA, Adapters, Prefix Tuning) and their applications in SE tasks.
To construct the SE task list used in our query, we additionally followed the taxonomy proposed by Watson \etal \cite{watson2022systematic} which provides a broad categorization of SE tasks studied in the context of DL4SE—an emerging area that applies deep learning techniques to solve a variety of software engineering problems.

To guide our retrieval process, we explicitly incorporated all 23 SE task names from Watson’s taxonomy into our search string. 
This design choice ensured broad task coverage and enabled the retrieval of papers relevant to both generative and non-generative SE tasks. 
Nevertheless, as discussed in \secref{subsec:filtering_process} and \secref{subsubsec:synthesis_approach}, several SE tasks from Watson’s taxonomy—such as Code Smells, Software Security, and Developer Forum Analysis—were not represented in our collection of papers.
In other words, not all the software engineering tasks have been explored in the context of PEFT.
A result that, although was expected, indicates that there is \emph{still} untapped potential to explore such optimization methods in broader SE domains. We present the results of our SLR in \secref{sec:results}.


We incorporated all SE task names reported in \tabref{tab:se_tasks}, which represent the tasks identified through our complete study selection, filtering, and snowballing process (\secref{subsec:filtering_process}). 

For Google Scholar, we used \textit{Publish or Perish (PoP)} software \cite{harzing2019publish}.
However, when using \textit{(PoP)} software to retrieve Google Scholar papers, the search bar has a 256-character limit. 
Due to this constraint, we split our search string into seven separate queries, each covering different subsets of PEFT techniques and SE tasks. 
This ensured that our search remained exhaustive while adhering to PoP's query limitations.
A few concrete examples are provided below--while all the queries can be retrieved from the replication package of our study \cite{replication}.

\begin{itemize}
    \item \texttt{('Parameter Efficient Fine Tuning' OR 'PEFT' OR 'LoRA' OR 'QLoRA' OR 'Adapters' OR 'Soft Prompts' OR 'Prefix Tuning' OR 'Prompt Tuning') AND ('Software Engineering' OR 'Automated Code Review' OR 'Static Analysis' OR 'Code Clone Detection' OR 'Bug Prediction' OR 'Commit Message Generation')}
    \item \texttt{('Parameter Efficient Fine Tuning' OR 'PEFT' OR 'LoRA' OR 'QLoRA' OR 'Adapters' OR 'Soft Prompts' OR 'Prefix Tuning' OR 'Prompt Tuning') AND ('Code Comprehension' OR 'Source Code Retrieval' OR 'Traceability' OR 'Bug-fixing Process' OR 'Code Smells')}
    \item \texttt{('Parameter Efficient Fine Tuning' OR 'PEFT' OR 'LoRA' OR 'QLoRA' OR 'Adapters' OR 'Soft Prompts' OR 'Prefix Tuning' OR 'Prompt Tuning') AND ('Software Testing' OR 'Non-code-related Software Artifacts' OR 'Software Energy Metrics' OR 'Program Synthesis')}
    
\end{itemize}



\subsection{Study Selection and Filtering Process} \label{subsec:filtering_process}

Building upon the initial filtering steps described in \secref{subsec:study_selection}, we now detail the exclusion and inclusion criteria used to finalize our set of primary papers. 
These criteria were applied to ensure that each study addressed both PEFT methods and SE tasks, and that it was published in top-tier venues within our defined time frame (2019-2025).

\subsubsection{Exclusion Criteria and Alignment Analysis}

Following our inclusion criteria, we applied a set of exclusion rules to filter out studies that fell outside the scope of our SLR.
Specifically, studies were excluded if they: (i) did not focus on SE-related tasks, (ii) were not published in one of the selected venues, or (iii) mentioned PEFT only as part of future work without presenting an actual implementation or evaluation.
After applying these criteria to the initial 1,146 retrieved records, we narrowed the selection down to \textbf{26 papers} that satisfied all requirements.


To ensure the reliability and quality of this selection, two independent reviewers (the first and the second author) conducted a detailed manual analysis of the 26 studies. 
Each reviewer systematically assessed the abstract, methodology, and contributions of the papers to verify whether they explicitly implemented, evaluated, or made a substantive contribution to PEFT techniques in the context of SE tasks.
During this review, \textbf{four disagreements} emerged two related to inclusion and two to exclusion decisions.
These cases were discussed collaboratively, and a consensus was reached to include all four disputed studies.
Due to access limitations, one paper \cite{LI2024112002} was unavailable for analysis. Therefore, our investigation proceeded with the remaining \textbf{25 papers} in the corpus--representing the outcome of Phase 3 of our SLR, also indicated with \figref{fig:slr_pipeline}--\circled{C}.

\subsubsection{Snowballing and Manual Addition of Studies} \label{subsubsec:snowballing}

After finalizing the initial set of papers, we applied backward snowballing to identify additional relevant studies that may have been missed during the database search. 
This involved reviewing the references of selected peer-reviewed papers, which led to identifying two additional arXiv studies that met our inclusion criteria. This stage of the SLR is depicted in \figref{fig:slr_pipeline}--\circled{D}.

We would like to remind the reader that, although these studies were not peer-reviewed at the time of inclusion, their presence in the reference lists of several reputable publications highlights their relevance and impact within the field.
We therefore included the work by Goel \etal~\cite{goel2022crossmodaltransfernaturallanguage}, cited by three related studies \cite{10173960, saberi2024utilizationpretrainedlanguagemodel, 10298587}, and the work by Ayupov \etal~\cite{ayupov2022parameter}, cited by four studies \cite{10.1145/3709358, chen-etal-2023-pass, saberi2024utilizationpretrainedlanguagemodel, 10298587}, both of which have been discussed or cited in multiple SE-focused PEFT papers \cite{10.1145/3709358, chen-etal-2023-pass, weyssow2023exploring}.

After incorporating these two additional studies, our final study set comprised \textbf{28 papers}. 
The distribution of these papers across publication venues is shown in \figref{fig:venue_selection}.

\subsection{Data Extraction and Synthesis} \label{subsec:data_extraction_synthesis}

Having finalized our corpus of 28 primary studies through rigorous search, filtering, and snowballing procedures (as described in \secref{subsec:study_selection} and \secref{subsec:filtering_process}), we proceeded to extract and organize data to answer our research questions. 
In the following subsections, we first describe our data extraction process and then outline the synthesis approach, including how we grouped software engineering tasks into generative and non-generative categories.

\subsubsection{Synthesis Approach} 
\label{subsubsec:synthesis_approach}

As introduced in \secref{sec:intro}, we aim to provide evidence regarding the use of PEFT optimization for large code models in both generative and non-generative SE tasks.
We define \textit{generative tasks} as activities in which models produce or transform code artifacts or generate related textual outputs—such as natural language summaries of code. These tasks typically involve generating sequences of tokens, which represent fundamental outputs that large code models produce for both natural language and code-related applications.
In contrast, \textit{non-generative tasks} focus on analyzing existing code by interpreting, classifying, retrieving, or detecting specific properties. Examples include Defect Detection, Code Clone Detection, Code Search, and Method Name Consistency Check.
This task classification serves as the foundation for our subsequent analysis, enabling us to map PEFT methods effectively to different types of software engineering tasks.

\subsubsection{Data Extraction}
\label{subsubsec:data_extract}

To systematically address our three research questions, we followed a structured extraction protocol that enabled systematic collection of relevant attributes from each of the 28 selected primary studies.

\textbf{RQ\textsubscript{1} (Software Engineering Tasks):}
For each study, we identified the primary SE tasks being addressed, categorizing them as either generative (\eg Code Generation, Code Summarization) or non-generative (\eg Defect Detection, Code Search). Where available, we also recorded high-level task attributes, such as input/output formats and application contexts, to characterize how PEFT methods were applied across different SE tasks. This classification facilitates subsequent analyses for RQ\textsubscript{2} and RQ\textsubscript{3}.

\textbf{RQ\textsubscript{2} (Model Architectures and PEFT Methods):}
We documented the base models used in each study, recording their architecture type (encoder-only, decoder-only, or encoder-decoder), model scale (\eg up to 70B parameters), and specific Transformer variants (\eg CodeT5-base, CodeLlama-13B). We also extracted information on PEFT methods employed in each SE task, including LoRA, QLoRA, Prefix Tuning, Adapters, Prompt Tuning, among others. Configuration-level details for commonly used PEFT methods were also collected, focusing on method-specific parameters and hyperparameters such as \texttt{rank} and the number of \texttt{target layers}, as well as general attributes, including the proportion of trainable parameters per task category.

\textbf{RQ\textsubscript{3} (Performance and Efficiency):}
We focused on comparing performance and computational efficiency between PEFT-optimized large code models and fully fine-tuned baseline models. For each study, we identified the best-performing PEFT methods (multiple methods if present) and recorded whether reported outcomes indicated improvements (\perfup \effup), equivalence (\perfneutral \effneutral), or declines (\perfdown \effdown) in performance and efficiency compared to the full fine-tuning baseline. Efficiency metrics such as reduced training time and memory consumption were noted where available. These findings were synthesized into a categorical summary, enabling task-wise comparisons of PEFT impacts on performance and efficiency.

Overall, our review encompasses 28 studies covering 19 distinct SE tasks, which we classify into \textcolor{black}{\textbf{twelve (12)}} generative and \textcolor{black}{\textbf{seven (7)}} non-generative task groups. These are further detailed in \tabref{tab:se_tasks}, where we report the task category in the leftmost column, followed by a full description of the software engineering task type--adopting the terminology established by Watson \etal \cite{watson2022systematic}. Finally, the third column reports the corresponding studies in which each task type was identified.

\begin{table}[h]
\centering
\small
\caption{Software Engineering Tasks Taxonomy into Generative and Non-Generative Categories with Corresponding Paper List.}
\label{tab:se_tasks}
\begin{tabular}{|>{\centering\arraybackslash}m{2cm}|>{\centering\arraybackslash}m{3.5cm}|>{\centering\arraybackslash}m{6cm}|>{\centering\arraybackslash}m{2.5cm}|}
\hline
\textbf{Category} & \textbf{SE Task} & \textbf{Task Description} & \textbf{Papers} \\
\hline
\multirow{24}{*}{\textbf{Generative}} 
& Automated Program Repair \textbf{(APR)} & Automatically generates patches to fix buggy code without human intervention. & \cite{li2024exploring, 10.1145/3691620.3695062, 10850625} \\
\cline{2-4}
& Code Generation \textbf{(CG)} & Generates source code from natural language descriptions or formal specifications. & \cite{10.1145/3714462, 10.1145/3637528.3671609, 10.1145/3639477.3639743, hajipour-etal-2024-simscood, choi-lee-2023-codeprompt, chen-etal-2023-pass, shi2023towards, ayupov2022parameter, yang2024cordacontextorienteddecompositionadaptation}, \textcolor{black}{\cite{wang2025beyond}} \\
\cline{2-4}
& Code Refinement \textbf{(CRF)} & Improves or restructures existing code while preserving its original functionality. & \cite{10.1145/3597503.3608132, hajipour-etal-2024-simscood, chen-etal-2023-pass, 10173960} \\
\cline{2-4}
& Code Review Generation \textbf{(CRG)} & Generates natural language review comments from code changes to provide immediate feedback during commits. & \cite{10.1145/3709358} \\
\cline{2-4}
& Code Summarization \textbf{(CSum)} & Generates concise natural language summaries describing code behavior or intent. & \cite{choi-lee-2023-codeprompt, 10.1145/3551349.3556909, 10173960, chen-etal-2023-pass, saberi2024utilizationpretrainedlanguagemodel, shi2023towards, 10298587, 10269066, ayupov2022parameter, 10.1145/3540250.3549113, wang2023one, afrin2025resource, 10.1145/3637528.3671609} \\
\cline{2-4}
& Code Translation \textbf{(CTrans)} & Converts code from one programming language to another while preserving functionality. & \cite{choi-lee-2023-codeprompt, chen-etal-2023-pass, 10298587, 10269066, ayupov2022parameter, 10.1145/3540250.3549113, 10.1145/3637528.3671609}, \textcolor{black}{\cite{wang2025beyond}} \\
\cline{2-4}
& Commit Message Generation \textbf{(CMG)} & Generates commit messages automatically from code changes to summarize their intent and content. & \cite{10589847, 10.1145/3709358} \\
\cline{2-4}
& Code Completion \textbf{(CComp)} & Predicts and completes the next code token or line based on the surrounding context. & \cite{shi2023towards, 10.1145/3637528.3671609}, \textcolor{black}{\cite{wang2025beyond}} \\
\cline{2-4}
& Just-in-time Comment Update \textbf{(JITCU)} & Automatically updates code comments in response to changes in the associated code. & \cite{10.1145/3709358} \\
\cline{2-4}
& Method Name Recommendation \textbf{(MNR)} & Suggests meaningful method names based on code content and contextual information. & \cite{10174153} \\
\cline{2-4}
& \textcolor{black}{Protocol Buffer Transformation \textbf{(PBT)}} & \textcolor{black}{Generates code functions that transform data structures between different format specifications.} & \textcolor{black}{\cite{wang2025beyond}} \\
\cline{2-4}
& Unit Test Generation \textbf{(UTG)} & Automatically generates unit test cases from source code to verify its correctness. & \cite{10.1145/3540250.3558959, 10.1145/3637528.3671609} \\
\hline
\multirow{13.5}{*}{\textbf{Non-Generative}} 
& Cloze Test \textbf{(CT)} & A probing task used to assess a model's understanding of code syntax without updating its parameters. & \cite{saberi2024utilizationpretrainedlanguagemodel, goel2022crossmodaltransfernaturallanguage} \\
\cline{2-4}
& Code Clone Detection \textbf{(CCD)} & Identifies syntactically or functionally similar code fragments to reduce redundancy and improve maintainability. & \cite{chen-etal-2023-pass, saberi2024utilizationpretrainedlanguagemodel, shi2023towards, 10298587, ayupov2022parameter, goel2022crossmodaltransfernaturallanguage} \\
\cline{2-4}
& Code Review \textbf{(CR)} & Assists the review process by recommending reviewers or identifying issues in code changes. & \cite{10.1145/3695993} \\
\cline{2-4}
& Code Search \textbf{(CS)} & Retrieves semantically relevant code snippets given a natural language query. & \cite{shi2023towards, 10269066, wang2023one} \\
\cline{2-4}
& Defect Detection \textbf{(DD)} & Predicts whether a code snippet contains bugs or security vulnerabilities. & \cite{chen-etal-2023-pass, 10298587, 10589847, 10269066, 10.1145/3540250.3549113} \\
\cline{2-4}
& \textcolor{black}{Header File Prediction \textbf{(HFP)}} & \textcolor{black}{Predicts the correct file paths for header dependencies given source file information.} & \textcolor{black}{\cite{wang2025beyond}} \\
\cline{2-4}
& Method Name Consistency Check \textbf{(MNCC)} & Determines whether a method name aligns with its implementation to detect naming inconsistencies. & \cite{10174153} \\
\hline
\end{tabular}
\end{table}

\begin{figure*}[h]
\centering
\includegraphics[width=\textwidth]{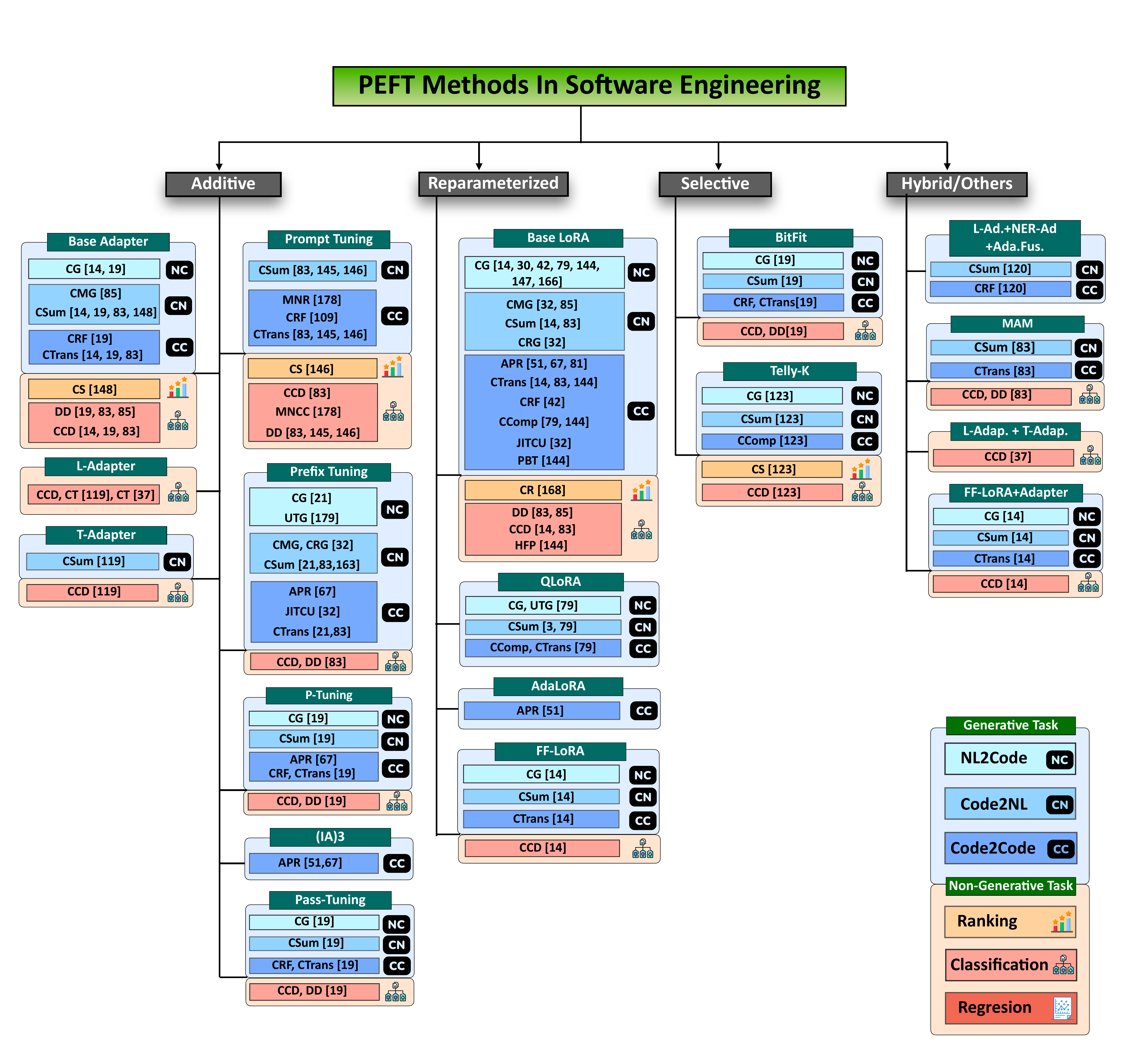}
\caption{PEFT Taxonomy for Software Engineering Tasks.}
\label{fig:peft_se_taxonomy} 
\end{figure*}

While \tabref{tab:se_tasks} outlines the range of software engineering tasks covered in our SLR, grouped by output type (generative vs. non-generative), \figref{fig:peft_se_taxonomy}--as anticipated in \secref{sec:intro} expands on this by presenting the taxonomy that visually  captures how PEFT methods are applied across task categories, offering a structured lens through which to assess current practices. By mapping adaptation techniques to task types, it highlights usage trends, gaps in coverage, and opportunities for more targeted or diversified applications of PEFT in software engineering.

\renewcommand{\arraystretch}{1.2} 
\renewcommand{\tabularxcolumn}[1]{>{\centering\arraybackslash}m{#1}} 

\section{Results}
\label{sec:results}

In this section, we present and analyze the findings of our investigation, structured according to the previously defined Research Questions. Each subsection corresponds to a specific RQ, offering a detailed examination of the results in relation to the objectives outlined in our study.\\

\subsection{RQ\textsubscript{1}: Which software engineering tasks have been explored using PEFT methods?}

This research question investigates and quantifies how PEFT methods are applied to enhance the efficiency, effectiveness, or both, of software engineering tasks.
From our analysis of 28 primary studies, we identified 19 distinct SE tasks where PEFT-based techniques have been applied. These tasks have already been categorized into \textit{generative} and \textit{non-generative} types in \secref{subsubsec:synthesis_approach}. 
\textcolor{black}{
Generative tasks typically involve creating or manipulating code-related artifacts as output (\eg source code, code documentation), often in an open-ended or token-by-token fashion. This includes tasks like code summarization, code generation, and commit message generation, even if the latter involves some form of templating--because the model must still synthesize content beyond merely selecting from fixed classes. On the other hand, non-generative tasks are those where the output space is bounded or categorical, such as classification, detection, or retrieval. For example, code classification--even though it produces an output--falls under the non-generative category because it selects from a finite set of labels and does not require constructing novel sequences of tokens.
We further organize the 12 generative tasks based on their input-output modalities as follows: \textbf{\textit{NL2Code}} tasks--which involve generating code from natural language, such as Code Generation and Unit Test Generation--appear in 12 studies; \textbf{\textit{Code2NL}} tasks--which involve translating code into natural language, such as Code Summarization, Commit Message Generation, and Code Review Generation--appear in 16 studies; and \textbf{\textit{Code2Code}} tasks--which involve code-to-code transformations or analysis, such as Code Translation, Code Refinement, Automated Program Repair, and others--are explored in 21 studies.
While generative tasks are categorized based on their input-output modalities due to the diversity and structural implications of the output (\ie text or code), non-generative tasks do not require such fine-grained modality breakdown. This is because the output across non-generative tasks is typically constrained to label spaces (\ie binary, multi-class, regression values).
To enhance clarity and better reflect the structural diversity of non-generative tasks, we classify them according to standard machine learning output paradigms \cite{zhang2023unifying, pena2025evaluating}. Specifically, we group them into the categories as follows -- \textit{Classification}: Tasks where the model predicts a discrete class label, such as Code Clone Detection, Defect Detection and Code Classification; \textit{Regression}: Tasks involving continuous-valued outputs; \textit{Ranking}: Tasks that return ordered lists or scores, such as Code Search or Traceability Link Recovery.
}


We obtained \textcolor{black}{17} papers where more than one SE task was supported via PEFT methods \cite{hajipour-etal-2024-simscood, choi-lee-2023-codeprompt, 10174153, chen-etal-2023-pass, 10173960, 10.1145/3709358, saberi2024utilizationpretrainedlanguagemodel, shi2023towards, 10589847, 10298587, 10269066, ayupov2022parameter, 10.1145/3540250.3549113, wang2023one, goel2022crossmodaltransfernaturallanguage, 10.1145/3637528.3671609, wang2025beyond}..

To ensure consistency in our classification, we clarify the following: when a paper applied a PEFT method to optimize a non-generative task that was used as part of a larger generative pipeline, we attributed the primary PEFT contribution to the non-generative task. This is because the PEFT technique was originally designed and evaluated for improving that specific component. The same principle applies in reverse: if a PEFT method was developed to enhance a generative task, even if it was later used within a broader classification or analysis framework, we still classified the contribution under the generative task.

\begin{figure*}[h]
\centering
\includegraphics[width=1\textwidth]{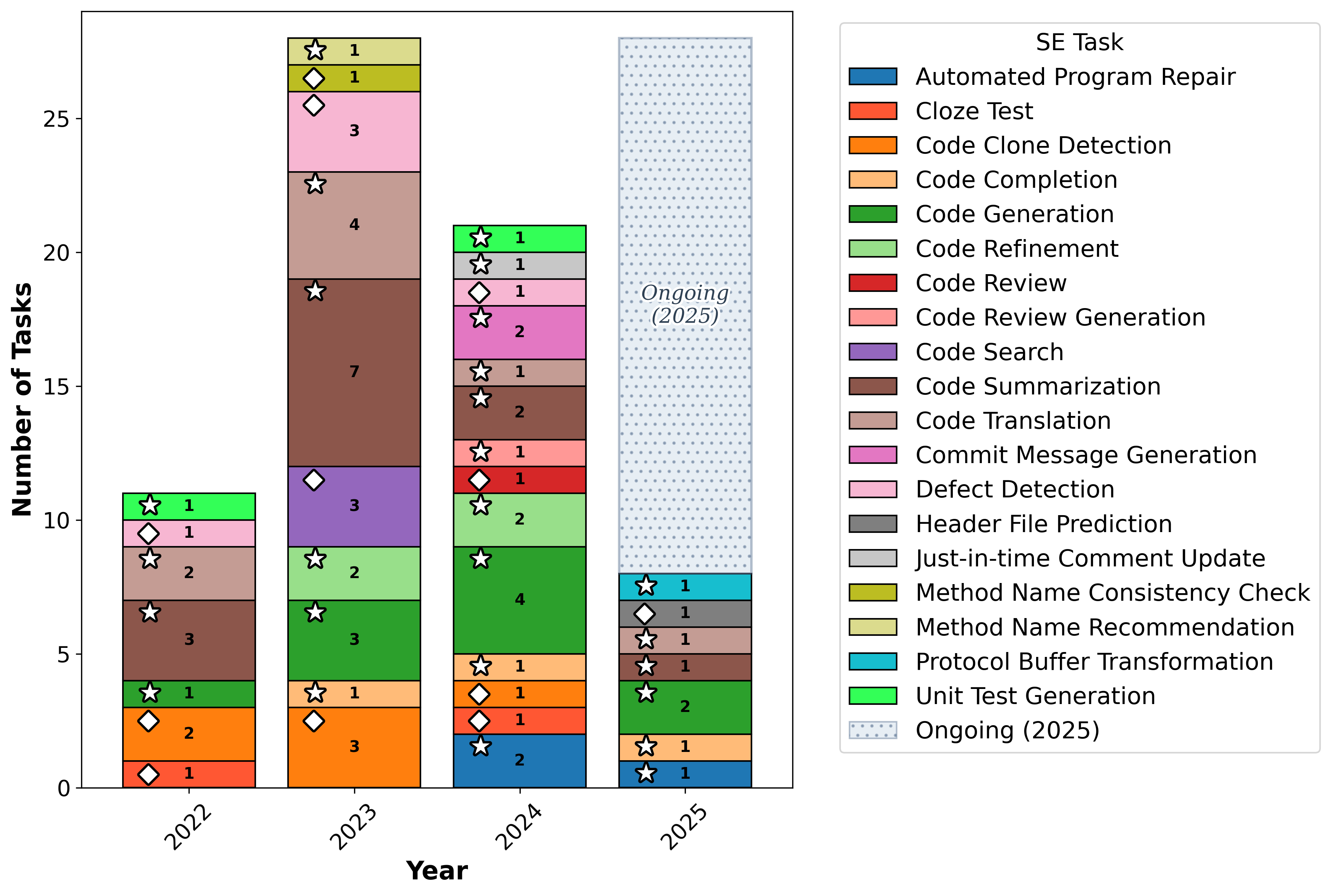}
\caption{
Trends of PEFT Techniques Applied to Software Engineering Tasks Over the Years. Generative tasks are marked with a star ($\star$), while non-generative tasks are marked with a diamond ($\diamond$). \textit{Note that 2025 is ongoing, and additional studies may still emerge throughout the year.}}
\label{fig:se_trend}
\end{figure*}

This approach allows us to build a taxonomy that accurately reflects the original goal and scope of each PEFT application. 
Between 2022 and 2025, research on PEFT in software engineering has grown significantly, with an increasing number of papers addressing diverse tasks. 
This growing trend is depicted in \figref{fig:se_trend} and highlights an early traction of Code Summarization (13 studies), and Code Clone Detection (6 studies), while Code Generation \textcolor{black}{(10 studies)} has only seen a notable rise in PEFT supports recently. Nonetheless, since our collection of relevant studies concluded on February 2025, we expect that additional papers-particularly in the areas of software engineering and broader applications-will be published by the end of the current year (2025).


Upon closer examination of generative tasks, we observe that Code Summarization accounted for $\sim$46.4\% (13 studies) and Code Generation with \textcolor{black}{10} studies ($\sim$35.7\%) are the most frequently explored tasks, reflecting strong interest in improving model efficiency for synthesis-based application.
Prefix Tuning and Prompt Tuning are among the earliest and most extensively tested methods for adapting Transformer-based models to Code Summarization with minimal trainable parameters. 
Adapter-based techniques are also popular, as evidenced in \cite{chen-etal-2023-pass,ayupov2022parameter,10298587,wang2023one}.

Furthermore, recent advancements in reparameterization-based methods (see \figref{fig:peft_se_taxonomy})--such as QLoRA—have shown promising results. For instance, Afrin \etal \cite{afrin2025resource} employed QLoRA to fine-tune large-scale code models, including CodeLlama-7B and DeepSeek-Coder-33B, on the CodeXGLUE benchmark \cite{lu:nips2021}, targeting the task of code summarization. The proposed approach attained performance comparable to that of fully fine-tuned counterparts, while demonstrating substantial gains in memory efficiency.

Ayupov \etal \cite{ayupov2022parameter} explores various PEFT methods like Adapter Tuning (AT), LoRA, FF-LoRA \cite{he2021towards}, and FF-LoRA + AT to fine-tune CodeT5-base and PLBART-base on the CodeSearchNet dataset \cite{husain:arxiv2019}, enabling parameter-efficient adaptation across diverse programming languages such as Python and Go--highlighting the interest from researchers of the field in exploring PEFT methods to cover SE-related activities that are developed to produce both type of information LCMs handle, \ie code and technical natural language.

To this extent, we want to highlight the usage of Reparametrized methods and particularly, LoRA for optimizing models primed for the production of code tokens. 
For example, Liu \etal \cite{10.1145/3637528.3671609} utilizes LoRA and QLoRA to fine-tune models such as QWen-14B and CodeLlama-Python-13B across benchmarks like HumanEval and MBPP, enabling efficient large-scale adaptation.

Not only have low-rank transformation methods been used for Code Generation, but additive approaches such as Prefix Tuning have also been explored. In this regard, Choi \etal \cite{choi-lee-2023-codeprompt} introduces an advanced Prefix Tuning method with input-dependent prompt templates and multi-word initialization to enhance CodeT5’s performance when generating Java code from natural language intents.

Moving forward, Code Translation also shows growing traction with \textcolor{black}{8} different papers exploring the extent to which PEFT techniques, including Additive, Reparametrized, as well as Hybrid, can serve the purpose of effective and efficient fine-tuning. For example, Liu \etal \cite{10298587} evaluates multiple PEFT methods including Adapter, LoRA, Prefix Tuning, Prompt Tuning, and Mix-And-Match (MAM) on CodeT5 and PLBART to translate code segments from  Java to C\#. In a different but related study, Wang \etal \cite{10269066} explore the capabilities of Prompt Tuning with varying sizes of CodeT5.

Within the domain of \cTc tasks, Code Refinement has emerged as a prominent area of research, with 4 studies investigating the advantages and limitations of applying PEFT techniques to the challenge of fixing buggy code. Peng \etal \cite{10.1145/3597503.3608132}, for example, examine the effectiveness of prompt-based tuning on a CodeT5-based model for bug-fixing tasks. Similarly, Saberi \etal \cite{10173960} employ both language-specific and task-specific adapters in combination with AdapterFusion to fine-tune CodeBERT, enabling it to generate meaningful fixes for buggy Java methods \cite{tufano2019empiricalstudylearningbugfixing}. In parallel, Chen \etal \cite{chen-etal-2023-pass} explore a range of PEFT techniques—including Pass-Tuning, BitFit, Adapters, and P-Tuning—built on top of CodeT5 and PLBART to correct faulty code components \cite{tufano2019empiricalstudylearningbugfixing}.

Similar findings are observed in the closely related area of Automated Program Repair, which includes three studies \cite{li2024exploring, 10.1145/3691620.3695062, 10850625} employing both Additive and Reparameterized PEFT methods. The Additive approaches include techniques such as (IA)$^3$, P-Tuning, and Prefix-Tuning, while the Reparameterized category features methods like AdaLoRA and BaseLoRA.
Notably, we identified that within each category, a specific PEFT variant—(IA)$^3$ for Additive methods and AdaLoRA for Reparameterized ones—was exclusively applied to APR tasks, suggesting a targeted use of these techniques for this particular application.


Additionally, we noted that tasks such as Code Review and Code Review Generation have received increasing attention from researchers \cite{10.1145/3709358, 10.1145/3695993, lu2023llama}. Through our SLR, we identified two distinct formulations of code review activities. The first, \textit{Code Review Generation}, is a generative task (as described in \tabref{tab:se_tasks}) that focuses on producing natural language feedback a reviewer would typically provide for a code change. The second, which we refer to as the \textit{Code Review}, belongs to the non-generative family and involves identifying or analyzing aspects of the code review process—such as classifying review comments, linking reviews to commits, or detecting review coverage patterns.
In this regard, Fan \etal \cite{10.1145/3709358} explored LoRA and Prefix Tuning on models such as InCoder and CodeGen, using a real-world diff-review dataset comprising over 138,000 instances. This highlights how PEFT methods can be effectively applied even in less frequently studied yet practically impactful SE tasks like Code Review Generation.

Other explored tasks include Just-in-time Comment Update, Method Name Recommendation, Code Completion, \textcolor{black}{Protocol Buffer Transformation} and Unit Test Generation.
Across these tasks, LoRA, Prefix Tuning, and Adapters remain the most frequently used methods, often applied to large-scale models such as CodeLlama \cite{roziere2023code}, CodeT5 \cite{wang2021codet5identifierawareunifiedpretrained}, and InCoder \cite{fried2022incoder}.

Similarly, in Commit Message Generation, Fan \etal \cite{10.1145/3709358} applies LoRA and Prefix Tuning to LLaMA and CodeGen models on the large-scale MCMD dataset by Tao \etal \cite{10.1007/s10664-022-10219-1}, which includes over 2 million commit-message pairs from top GitHub repositories. 

When PEFT methods are applied to fine-tune large code models for method name recommendation, compelling evidence is provided by Zhu \etal \cite{10174153}. In their study, Prompt Tuning is initialized using CodeT5 and enhanced by incorporating contextual code elements. The authors reframed the method name consistency check as a classification problem, leading to improved naming accuracy. This example illustrates how PEFT techniques can enable efficient and targeted adaptation, even for less frequently explored software engineering tasks.

Moving beyond generative tasks, PEFT techniques have also been explored for non-generative SE applications, with Code Clone Detection being the most prominent among them, appearing in 6 studies ($\sim$21.4\%). 
\textcolor{black}{Analyzing the distribution of non-generative tasks across machine learning paradigms reveals clear patterns. Classification tasks dominate, with 16 studies covering Cloze Test (2), Code Clone Detection (6), Code Review (1), Defect Detection (5), Method Name Consistency Check (1), and Header File Prediction (1). Ranking tasks appear in only 3 studies, all in Code Search applications, and no regression-based tasks were identified. Overall, approximately $\sim$84\% of non-generative PEFT applications focus on discrete classification problems, split into 8 papers on binary classification ($\sim$66.7\%), 2 on multi-class classification ($\sim$16.7\%), and 2 addressing both ($\sim$16.7\%).}
To this end, a relevant approach is Telly-K, proposed by Shi \etal \cite{shi2023towards}, which selectively tunes the bottom layers of UniXcoder and GraphCodeBERT to detect Java code clones on the BigCloneBench dataset \cite{svajlenko2016bigcloneeval} which is Defect Detection is another important non-generative task where PEFT methods have shown promise, appearing in 5 studies. 
This task focuses on identifying buggy code, often using C or Java datasets like Devign \cite{zhou2019devigneffectivevulnerabilityidentification} and JIT-Defects4J. 
For example, Liu \etal \cite{10589847} explores Adapter Tuning and LoRA on a diverse set of models—including CodeBERT, GraphCodeBERT, and PLBART using the JIT-Defects4J dataset with over 16K training instances to detect Just-in-time Software Defects. 
Similarly, Wang \etal \cite{10269066} applies Prompt Tuning to CodeBERT and CodeT5 variants on the Devign dataset, which includes over 27K C code snippets with nearly half labeled as defective. These efforts highlight how PEFT enables effective adaptation for real-world bug detection scenarios with limited parameter updates.

It is worth noting that Code Search appears in 3 studies, primarily evaluated on the CodeSearchNet dataset \cite{husain:arxiv2019}--while for Cloze Test, which appears in 2 studies, we find applications of L-Adapter \cite{goel2022crossmodaltransfernaturallanguage} trained on 
CodeSearchNet and CodeNet \cite{puri2021codenet}. Method Name Consistency check (1 study) is addressed by Zhu \etal \cite{10174153} using Prompt Tuning with CodeT5, where contextual prompts are crafted to convert the problem into a classification task, effectively enhancing consistency evaluation. 

\vspace{10pt}

In summary, our analysis reveals a growing and diverse adoption of PEFT techniques across the landscape of software engineering tasks, encompassing both generative tasks (\eg Code Summarization, Code Generation, Code Translation) and non-generative tasks (\eg Defect Detection, Code Clone Detection, and Code Search). This widespread applicability highlights PEFT’s flexibility in adapting large code models with minimal parameter updates, making them more accessible for practical use in real-world software engineering scenarios where computational constraints are common.

\boxwithsideicon{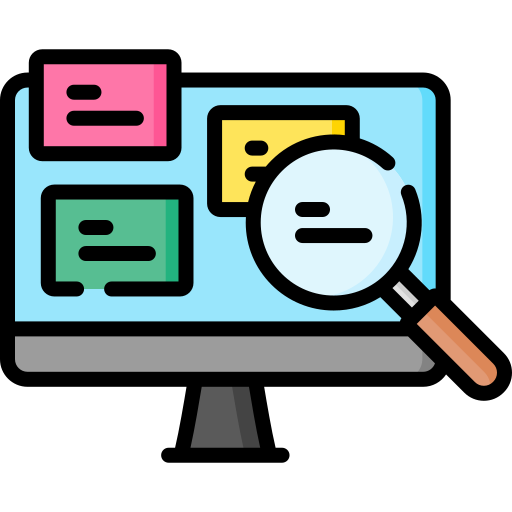}{Key Findings (RQ\textsubscript{1})}{%
PEFT techniques have been applied to \textcolor{black}{19 diverse SE tasks}, with a strong focus on generative tasks and growing interest in non-generative ones. While tasks like Code Summarization and Generation remain the most studied, several emerging tasks have also seen initial exploration. Common PEFT methods include Prompt Tuning, Adapter, LoRA, (IA)$^3$, and QLoRA, applied across popular models such as CodeT5, CodeLlama, and CodeBERT. These findings demonstrate the adaptability and effectiveness of PEFT in software engineering with different model architectures and varying sizes. 
\textcolor{black}{
Task concentration patterns reveal that Code Summarization represents $\sim$46.4\% of studies while Code Generation accounts for $\sim$35.7\%, indicating clear research priorities in synthesis-based applications. Cross-task versatility analysis shows that $\sim$60.7\% of studies apply PEFT to multiple SE tasks, demonstrating broader methodological flexibility than single-task approaches. The distribution across task categories shows non-generative tasks like Code Clone Detection representing $\sim$21.4\% of studies, revealing systematic coverage of both generative and non-generative software engineering activities.
}
}

\vspace{0.2cm}

\boxwithsideicon{images/future.png}{Future Directions (RQ\textsubscript{1})}{%
Future research should expand PEFT applications to underexplored and emerging software engineering tasks that remain less represented in current literature. This includes areas such as testing, documentation, and code quality analysis, as well as forward-looking tasks like API Recommendation and Automated Refactoring. Another promising direction is cross-task adaptation, where a single PEFT configuration is designed to generalize across multiple activities. Additionally, investigating task-specific PEFT variants can offer insights into which methods work best for particular task types. Such efforts would help bridge research innovations with practical software development needs.%
}

\subsection{RQ\textsubscript{2}: Which deep learning models and architectures are commonly optimized with PEFT methods?} \label{sec:peft_methods}

In RQ$_{1}$, we examined the range of software engineering tasks that have been addressed using PEFT methods, highlighting trends over time. 
Building on that foundation, this section shifts focus to the deep learning models and particularly Transformers architectures \cite{vaswani:nips2017} optimized with PEFT techniques across various software engineering tasks. 
In particular, we analyze the types of Transformer models that have been optimized using PEFT and their alignment with different development activities, focusing on whether certain architectures (\eg encoder-only, decoder-only, encoder-decoder) are more commonly employed for PEFT-optimization in comparison to others. 
Furthermore, we examine how various PEFT methods (\eg LoRA, QLoRA, Prefix Tuning, Adapters) are applied across different model architectures, and whether certain learning algorithms or optimization strategies consistently achieve better results. Ultimately, by analyzing these factors, we aim to identify patterns that illustrate how PEFT techniques are shaping the adaptation of deep learning models—particularly Transformer-based architectures—for software engineering applications.

\subsubsection{Base Models and Architectures in PEFT Software Engineering Optimization.}

Before detailing the results of this analysis, we remind the reader that one paper can employ more than one techniques, as a results of that, we found that encoder-decoder architectures are the most prevalent in PEFT-based software engineering optimization, appearing in 29 software engineering task instances across 12 studies \cite{hajipour-etal-2024-simscood, choi-lee-2023-codeprompt, 10174153, 10.1145/3551349.3556909, chen-etal-2023-pass, 10298587, 10269066, ayupov2022parameter, 10.1145/3540250.3549113, wang2023one, 10.1145/3597503.3608132, 10.1145/3540250.3558959}. This is followed by encoder-only architectures with 22 applications across 10 studies \cite{hajipour-etal-2024-simscood, 10173960, saberi2024utilizationpretrainedlanguagemodel, shi2023towards, 10589847, 10298587, 10269066, 10.1145/3540250.3549113, wang2023one, goel2022crossmodaltransfernaturallanguage}, while we found that auto-regressive models (\ie decoder-only architectures) are employed to automate \textcolor{black}{19 software engineering activities across 12 studies} \cite{li2024exploring, 10.1145/3637528.3671609, 10.1145/3639477.3639743, yang2024cordacontextorienteddecompositionadaptation, hajipour-etal-2024-simscood, 10.1145/3709358, 10.1145/3695993, 10.1145/3691620.3695062, afrin2025resource, 10850625, 10.1145/3714462, wang2025beyond}. These findings indicate a clear preference for encoder-decoder architectures among researchers, likely due to their effectiveness in handling both understanding and generation tasks common in SE applications. \textcolor{black}{
It is important to note that our goal in examining model architectures is not to suggest that architecture type prescriptively determines the choice of PEFT method. Instead, this analysis aims to document how current practices reveal consistent patterns in how PEFT techniques are applied across model families. By empirically mapping the distribution of PEFT methods over different architectures, we capture how practical considerations--such as model structure, parameter scale, and task orientation--correlate with the adoption of particular strategies.
    }

When looking at the base models, our analysis reveals that CodeT5 is the most frequently adapted LCMs, appearing in a total of 19 times among all SE studies across 8 studies \cite{hajipour-etal-2024-simscood, choi-lee-2023-codeprompt, 10174153, chen-etal-2023-pass, 10589847, 10298587, wang2023one, 10.1145/3597503.3608132}. 

For the encoder-only model GraphCodeBERT, we report 10 software engineering task instances across 5 studies \cite{hajipour-etal-2024-simscood, saberi2024utilizationpretrainedlanguagemodel, shi2023towards, 10589847, 10298587}. These tasks primarily fall into the \cTc and \cTn categories. 

Moving forward, CodeBERT--an encoder-only transformer model, has been used in 9 software engineering task instances across 6 studies \cite{10173960, saberi2024utilizationpretrainedlanguagemodel, 10589847, 10298587, 10269066, 10.1145/3540250.3549113}, while PLBART appears in 9 SE task instances across 6 studies \cite{10173960, saberi2024utilizationpretrainedlanguagemodel, 10589847, 10298587, 10269066, 10.1145/3540250.3549113}, and UniXcoder is applied to 8 tasks across 3 studies \cite{shi2023towards, 10589847, wang2023one}. 

When examining specific model variants, the CodeLlama family shows particularly flexible adoption, with CodeLlama-7B appearing in 7 SE instances across 5 studies \cite{li2024exploring, 10.1145/3639477.3639743, 10.1145/3709358, 10.1145/3691620.3695062, afrin2025resource}, \textcolor{black}{CodeLlama-13B in 6 SE instances across 3 studies \cite{li2024exploring, 10.1145/3709358, wang2025beyond}, and CodeLlama-34B in 5 studies \cite{10.1145/3637528.3671609, 10.1145/3639477.3639743, afrin2025resource, 10850625, wang2025beyond}}. Crossing tasks and models, we find out that CodeLlama-7B is used across diverse tasks, including Automated Program Repair, Code Generation, Commit Message Generation, and Code Summarization. CodeLlama-13B extends this usage to \textcolor{black}{Code Generation, Code Completion, Code Translation, and Unit Test Generation, Protocol Buffer Transformation}, while CodeLlama-34B supports more resource intensive tasks such as Code Completion and Automated Program Repair. This shows that the CodeLlama models are effective across a wide range of tasks, including \nTc tasks like Code Generation, \cTn tasks like Code Summarization and Code Review Generation.

This distribution highlights researchers' strategic selection of models that balance generation capability with efficiency, such as training time and memory usage, leading to a growing emphasis on larger, specialized models like CodeLlama. A full analysis of these trade-offs—considering both functional performance and computational efficiency-is provided in \secref{subsec:model-analysis}. 

Several recent studies have explored a range of model architectures in their evaluations, including StarCoder-15B \cite{10850625,10.1145/3637528.3671609}, DeepSeek-Coder variants \cite{li2024exploring,10.1145/3637528.3671609,afrin2025resource, wang2025beyond}, and QWen-14B \cite{10.1145/3637528.3671609}, reflecting interest in evaluating both general purpose language models and those specialized for code related tasks.
The architecture distribution reveals clear task-specific patterns. 
Encoder-decoder models (CodeT5, PLBART) maintain their dominance in generation-heavy tasks, and thus, they are naturally applied in the context of generative applications. On the other hand, encoder-only models (CodeBERT, GraphCodeBERT) remain preferred for analysis-oriented tasks. 
Finally, decoder-only models (particularly CodeLlama variants) are demonstrating growing adoption across both Code Generation and Automated Program Repair tasks, as evidenced by multiple studies \cite{li2024exploring,10.1145/3691620.3695062,10.1145/3714462,10.1145/3637528.3671609,10.1145/3639477.3639743,hajipour-etal-2024-simscood,10.1145/3709358}. 
Decoder-only models appeared exclusively in generative tasks. Specifically, they are used in \nTc tasks such as Code Generation \textcolor{black}{(5 studies)}, \cTn tasks such as Code Summarization, Commit Message Generation, and Code Review Generation (each reported in 1 study), and \cTc tasks such as Automated Program Repair (2 studies), Code Translation, Code Refinement, \textcolor{black}{Protocol Buffer Transformation} and Just-in-Time Comment Update (each reported in 1 study).
The evolution of model selection shows a marked progression from earlier focus on established architectures to current exploration of larger, more specialized models. 
This shift is particularly evident in the rising adoption of 13B+ parameter models like CodeLlama variants (7B/13B/34B), DeepSeek-Coder-33 \cite{10.1145/3637528.3671609,afrin2025resource, wang2025beyond}, and LLaMA-2-70B \cite{10.1145/3637528.3671609}, enabled by advancing PEFT techniques that make such models more accessible even in resource-constrained environments.

\subsubsection{PEFT Methods Across Software Engineering Tasks and Model Architectures} \label{subsec:peft_methods_on_SE}

In this section, we explore how PEFT methods are distributed across various software engineering tasks, distinguishing them between generative and non-generative. Among the identified techniques, Base LoRA is the most dominant, appearing in \textcolor{black}{29} SE tasks. The Adapter family has been applied across 23 software engineering tasks, encompassing specialized variants such as L-Adapters and T-Adapters. Other frequently used techniques include Prompt Tuning (14 studies), Prefix Tuning (13), and P-Tuning (7), which leverage soft or continuous prompts to inject task-specific knowledge into frozen pre-trained models. Collectively, these methods demonstrate significant effectiveness in reducing training costs while maintaining or enhancing performance, affirming their value in scalable and efficient model adaptation across a diverse set of software engineering tasks.

For generative tasks, we observe distinct patterns in the application of PEFT methods. 
In Automated Program Repair, three studies adopted LoRA \cite{li2024exploring, 10.1145/3691620.3695062, 10850625}, highlighting an interesting correlation task-method.
Additionally, one study \cite{li2024exploring} explored other PEFT strategies such as P-Tuning and Prefix Tuning independently, while another study evaluated AdaLoRA and (IA)$^3$ for APR \cite{10850625}. 
This reflects a diverse experimentation landscape where different lightweight fine-tuning techniques are applied across studies to assess their effectiveness.

\textcolor{black}{Code Generation is addressed in 10 studies, making it one of the most frequently explored generative SE tasks in the context of PEFT. LoRA remains the dominant technique, appearing in 7 studies}~\cite{10.1145/3714462, 10.1145/3637528.3671609, 10.1145/3639477.3639743, hajipour-etal-2024-simscood, ayupov2022parameter, yang2024cordacontextorienteddecompositionadaptation, wang2025beyond}, including combinations of FF-LoRA variants with Adapter Tuning, one study implemented Code Summarization, Code Clone Detection, Code Generation, and Code Translation \cite{ayupov2022parameter}. 

Despite the wide employment of Reparametrized techniques, it is important to acknowledge the community’s commitment to exploring the full spectrum of PEFT methods—ranging from Additive to Hybrid approaches—for Code Generation tasks. This growing body of work reflects a clear interest in identifying which PEFT strategies most effectively adapt large models to one of the most central and computationally demanding software engineering activities.

A further application of SE automation revolves around the generation of unit tests--referred to as Unit Test Generation in our taxonomy \figref{fig:peft_se_taxonomy}. In particular, for such a task we have found two distinct studies, each employing distinct PEFT approaches. Liu \etal \cite{10.1145/3637528.3671609} applied QLoRA within a multi-task fine-tuning framework (MFT-5Tasks), while Zlotchevski \etal \cite{10.1145/3540250.3558959} employed Prefix Tuning, wherein 200 trainable virtual tokens were prepended to a frozen BART model.



Looking at SE tasks for which code changes optimization is requested, we find a surprisingly widespread application of PEFT techniques to support Code Refinement activities (\ie bug-fixing).  This peculiar task, has been explored through a diverse range of PEFT techniques, which—consistent with findings from Code Generation—span the full spectrum of parameter-efficient optimization strategies. In this context, Nuo \etal~\cite{chen-etal-2023-pass} introduced Pass-Tuning, a lightweight method that tunes less than 1\% of parameters by leveraging U-AST (Upgraded Abstract Syntax Tree) token sequences within a graph-attentional model. Building on structure-aware prompting, Yun \etal~\cite{10.1145/3597503.3608132} applied Prompt Tuning to CodeT5-base, crafting domain-specific prompts based on AST-derived fix templates. 
Saberi \etal~\cite{10173960} proposed a fusion-based approach, combining a Java-specific L-Adapter with a token-type-aware NER-Adapter via AdapterFusion to improve bug-fixing effectiveness. Similarly, Hajipour \etal~\cite{hajipour-etal-2024-simscood} investigated the use of LoRA as a parameter-efficient alternative for Code Refinement tasks. Traditional Prompt Tuning appears in one study~\cite{10.1145/3597503.3608132}, while a multi-perspective adaptation strategy—incorporating L-Adapter, NER-Adapter, and AdapterFusion—is explored in a different research~\cite{10173960}. Additionally, one study applies a comprehensive suite of PEFT methods—including Pass-Tuning, BitFit, Adapters, and P-Tuning—to enhance source code refinement~\cite{chen-etal-2023-pass}. 

Still in the domain of \cTc tasks, Code Translation is addressed in \textcolor{black}{8 studies}~\cite{choi-lee-2023-codeprompt, chen-etal-2023-pass, 10298587, 10269066, ayupov2022parameter, 10.1145/3540250.3549113, 10.1145/3637528.3671609, wang2025beyond}, \textcolor{black}{while Code Completion is explored in 3 studies}~\cite{shi2023towards, 10.1145/3637528.3671609, wang2025beyond}. Notably, for the latter, Shi \etal~\cite{shi2023towards} introduce a customized PEFT technique called \textit{Telly-K}, which modifies the training procedure by selectively freezing the lower layers of the model. This approach is designed to preserve foundational representations while reducing computational overhead during fine-tuning. Beyond these, several other PEFT techniques have been applied to \cTc tasks, where code manipulation is central. \textit{Prefix Tuning} is adopted in 2 studies~\cite{choi-lee-2023-codeprompt, 10298587}, while Prompt Tuning is utilized in 3 studies~\cite{10298587, 10269066, 10.1145/3540250.3549113}. 

Reparametrized methods such as LoRA and its variants—including FF-LoRA and FF-LoRA + AT—are employed in 2 further studies~\cite{10298587, ayupov2022parameter}. Additionally, Adapter-based strategies such as MAM, as well as lightweight approaches like Pass-Tuning, BitFit, and P-Tuning, are explored individually across different works~\cite{chen-etal-2023-pass, 10298587, ayupov2022parameter}. 

In light of this analysis, a noteworthy trend emerges: the wide range of PEFT methods outlined above reflects the community’s active exploration of diverse strategies to determine the most effective parameter-efficient techniques for code transformation tasks. This ongoing experimentation highlights both the complexity of these tasks and the growing interest in tailoring fine-tuning approaches to their specific requirements.

Turning to \cTn tasks, Code Review Generation is investigated in 1 study~\cite{10.1145/3709358}, which independently applies both LoRA and Prefix Tuning for efficient adaptation. Code Summarization, however, emerges as the most frequently studied \cTn task, appearing in 13 studies. A broad set of PEFT techniques has been explored in this context. Among additive methods, Prefix Tuning and Adapter-based approaches—including L-Adapters, T-Adapters, and AdapterFusion—are used in 3~\cite{choi-lee-2023-codeprompt, 10.1145/3551349.3556909, 10298587} and 6 studies~\cite{10173960, chen-etal-2023-pass, saberi2024utilizationpretrainedlanguagemodel, 10298587, ayupov2022parameter, wang2023one}, respectively. Prompt Tuning is featured in 3 studies~\cite{10298587, 10269066, 10.1145/3540250.3549113}, while LoRA and its variants (FF-LoRA and \textit{FF-LoRA + AT}) are applied in 2 studies~\cite{10298587, ayupov2022parameter}. Additional techniques such as Pass-Tuning, BitFit, P-Tuning, and MAM are also explored. 

Notably, QLoRA is also employed in two \cTn studies~\cite{10.1145/3637528.3671609, afrin2025resource}. In particular, Afrin \etal~\cite{afrin2025resource} successfully applied QLoRA to fine-tune large language models with up to 4 billion parameters for the task of Code Summarization, demonstrating the feasibility of efficient adaptation at varying scales.

Commit Message Generation follows with 2 studies--both by relying on the application of LoRA~\cite{10.1145/3709358, 10589847}. However, while study combines LoRA with Prefix Tuning~\cite{10.1145/3709358}, the other integrates it with Adapter Tuning~\cite{10589847}.


Non-generative SE tasks typically adopt more targeted adaptation strategies, leveraging structured PEFT methods like Adapter and Prompt Tuning to enhance code understanding and classification. Cloze Test for example, is covered by 2 studies \cite{saberi2024utilizationpretrainedlanguagemodel, goel2022crossmodaltransfernaturallanguage}, both employing Adapter-based methods. 

Code Clone Detection has been explored across 6 studies, using a variety of PEFT methods, particularly, Adapter-based strategies. Study by Goel \etal \cite{goel2022crossmodaltransfernaturallanguage} employs L-Adapter and T-Adapter modules trained on datasets such as Big Clone Bench (BCB), POJ-104, and SCD-88 to perform Code Clone Detection. Other studies \cite{chen-etal-2023-pass, 10298587, ayupov2022parameter} adopt modular PEFT strategies including Adapter, P-Tuning, Pass-Tuning, and BitFit. 

Code Search is addressed in 3 studies, each applying distinct PEFT methods. A work by Wang \etal \cite{10269066}, Prompt Tuning was applied using CodeBERT in a two-tower architecture, where hard prompts were designed separately for code and natural language inputs to generate embedding vectors for cosine similarity based retrieval. A study by Wang \etal \cite{wang2023one} use Adapter with UniXcoder, by configuring lightweight 5M-parameter setup that outperformed fully fine-tuned baselines such as CodeBERT and PLBART. Another study \cite{shi2023towards} explored PEFT methods alongside traditional architectures but is excluded here from detailed discussion, as its method (Telly-K) was previously covered. These works highlight the applicability of Prompt and Adapter-based strategies for retrieval tasks without adjusting the whole set of parameters describing the underlying model.

In the domain of non-generative tasks, Defect Detection has been explored in 5 studies, showcasing a diverse range of PEFT techniques aimed at identifying faulty or vulnerable code segments. Among these, additive methods appear to be the preferred choice, with Prompt Tuning adopted in 3 studies~\cite{10269066, 10.1145/3540250.3549113, 10298587}, underscoring its effectiveness for classification-oriented tasks.

It is also worth highlighting that several studies implement multiple PEFT techniques within the same experimental setup—including Adapters, LoRA, and others~\cite{10589847, 10298587}—although these are typically applied independently rather than in an integrated manner. Notably, Chen \etal~\cite{chen-etal-2023-pass} evaluate a recurring combination of methods, including Pass-Tuning, BitFit, Adapter, and P-Tuning, which have also been observed in other software engineering tasks, suggesting their broad applicability across different adaptation scenarios.

\medskip

Overall, while non-generative tasks employ a targeted subset of PEFT techniques, generative tasks display a broader range of strategies—with particularly high adoption of LoRA and prompt-based methods. This trend reflects the growing need for scalable, low-resource generative models in software engineering, supporting the increasing integration of generative AI across a wide array of development activities.

\tabref{tab:peft_distribution} summarizes the distribution of PEFT methods across generative and non-generative software engineering tasks, highlighting which specific papers employ each method. Software Engineering task abbreviations are defined in \tabref{tab:se_tasks}.

\begin{table}[htbp]
    \centering
    \caption{PEFT Methods and Sub-Variants Across Software Engineering Tasks (Generative vs. Non-Generative). \textit{Note: Multiple SE tasks followed by a single citation indicate that all tasks were addressed in the same paper.}}
    \label{tab:peft_distribution}
    \renewcommand{\arraystretch}{1.3}
    
    \setlength{\tabcolsep}{4pt}
    \begin{tabularx}{\textwidth}{|>{\centering\arraybackslash\hsize=0.8\hsize}X|
                               >{\centering\arraybackslash\hsize=1.1\hsize}X|
                               >{\centering\arraybackslash\hsize=1.05\hsize}X|
                               >{\centering\arraybackslash\hsize=1.05\hsize}X|}
        \hline
        \textbf{Main Method} & \textbf{Sub-Method} & \textbf{Generative Tasks} & \textbf{Non-Generative Tasks} \\
        \hline
        \multirowcell{-1}[-23pt]{LoRA-based} 
            & Base LoRA & APR\cite{li2024exploring,10.1145/3691620.3695062,10850625}, CComp\cite{10.1145/3637528.3671609} \textcolor{black}{\cite{wang2025beyond}}, CG\cite{10.1145/3639477.3639743,hajipour-etal-2024-simscood,ayupov2022parameter,10.1145/3714462, yang2024cordacontextorienteddecompositionadaptation, 10.1145/3637528.3671609}, \textcolor{black}{\cite{wang2025beyond}} CMG\cite{10.1145/3709358,10589847}, CRF\cite{hajipour-etal-2024-simscood}, CRG\cite{10.1145/3709358}, CSum\cite{10298587,ayupov2022parameter}, CTrans\cite{10298587,ayupov2022parameter} \textcolor{black}{\cite{wang2025beyond}}, JITCU\cite{10.1145/3709358}, \textcolor{black}{PBT\cite{wang2025beyond}} & CCD\cite{10298587,ayupov2022parameter}, CR\cite{10.1145/3695993}, DD\cite{10589847,10298587}, \textcolor{black}{HFP\cite{wang2025beyond}} \\
            \cline{2-4}
            & QLoRA & CComp, CTrans\cite{10.1145/3637528.3671609}; CG, UTG \cite{10.1145/3637528.3671609}; CSum\cite{afrin2025resource, 10.1145/3637528.3671609} & $\times$ \\
            \cline{2-4}
            & AdaLoRA & APR\cite{10850625} & $\times$ \\
            \cline{2-4}
            & FF-LoRA & CG, CSum, CTrans\cite{ayupov2022parameter} & CCD\cite{ayupov2022parameter} \\
        \hline
        \multirowcell{1.2}[-14pt]{Adapter-based}
            & Base Adapter & CG\cite{chen-etal-2023-pass,ayupov2022parameter}, CMG\cite{10589847}, CRF\cite{chen-etal-2023-pass}, CSum\cite{chen-etal-2023-pass,10298587,ayupov2022parameter,wang2023one}, CTrans\cite{chen-etal-2023-pass,10298587,ayupov2022parameter} & CCD\cite{chen-etal-2023-pass,10298587,ayupov2022parameter}, CS\cite{wang2023one}, DD\cite{chen-etal-2023-pass,10589847,10298587} \\
            \cline{2-4}
            & L-Adapter & $\times$ & CCD, CT\cite{saberi2024utilizationpretrainedlanguagemodel}, CT\cite{goel2022crossmodaltransfernaturallanguage} \\
            \cline{2-4}
            & T-Adapter & CSum\cite{saberi2024utilizationpretrainedlanguagemodel} & CCD\cite{saberi2024utilizationpretrainedlanguagemodel} \\
        \hline
        \multirowcell{3.7}[-32pt]{Prompt-based}
            & Prompt Tuning & CRF\cite{10.1145/3597503.3608132}, CSum, CTrans \cite{10269066}, CSum, CTrans \cite{10298587}, CSum, CTrans \cite{10.1145/3540250.3549113}, MNR\cite{10174153} & CCD\cite{10298587}, CS\cite{10269066}, DD\cite{10298587,10269066,10.1145/3540250.3549113}, MNCC\cite{10174153} \\
            \cline{2-4}
            & Prefix Tuning & APR\cite{li2024exploring}, CG\cite{choi-lee-2023-codeprompt}, CMG, CRG\cite{10.1145/3709358}, CSum\cite{10.1145/3551349.3556909,10298587,choi-lee-2023-codeprompt}, CTrans\cite{10298587,choi-lee-2023-codeprompt}, JITCU\cite{10.1145/3709358}, UTG\cite{10.1145/3540250.3558959} & CCD, DD\cite{10298587} \\
            \cline{2-4}
            & P-Tuning & APR\cite{li2024exploring}, CG, CRF, CSum, CTrans\cite{chen-etal-2023-pass} & CCD, DD\cite{chen-etal-2023-pass} \\
            \cline{2-4}
            & Pass-Tuning & CG, CRF, CSum, CTrans\cite{chen-etal-2023-pass} & CCD, DD\cite{chen-etal-2023-pass} \\
            \cline{2-4}
            & (IA)$^3$ & APR\cite{li2024exploring,10850625} & $\times$ \\
        \hline
        \multirowcell{6.5}[-0pt]{Hybrid/Others}
            & BitFit & CG, CRF, CSum, CTrans\cite{chen-etal-2023-pass} & CCD, DD\cite{chen-etal-2023-pass} \\
            \cline{2-4}
            & FF-LoRA + AT & CG, CSum, CTrans\cite{ayupov2022parameter} & CCD\cite{ayupov2022parameter} \\
            \cline{2-4}
            & L-Adapter + T-Adapter & $\times$ & CCD\cite{goel2022crossmodaltransfernaturallanguage} \\
            \cline{2-4}
            & \makecell{L-Adapter + NER-Adapter \\ + AdapterFusion} & CRF, CSum\cite{10173960} & $\times$ \\
            \cline{2-4}
            & Telly-K & CComp, CG, CSum\cite{shi2023towards} & CCD, CS\cite{shi2023towards} \\
            \cline{2-4}
            & MAM & CSum, CTrans\cite{10298587} & CCD, DD\cite{10298587} \\
        \hline
    \end{tabularx}
\end{table}

\begin{figure*}[h]
\centering
\includegraphics[width=\textwidth]{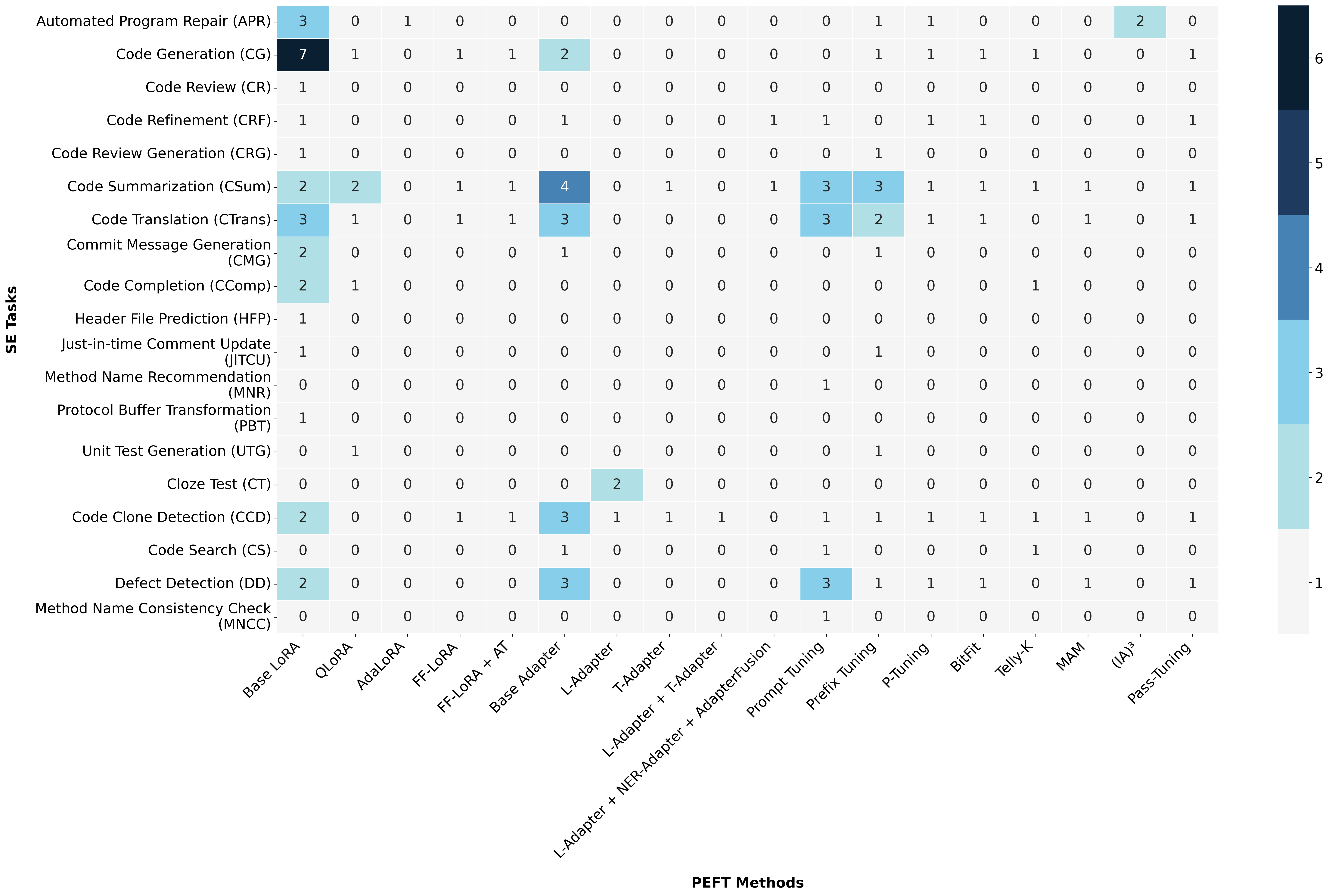}
\caption{Heatmap of PEFT Methods Across Software Engineering Tasks. The numbers represent the count of studies that utilized a
particular PEFT method for a given task.}
\label{fig:heatmap_peft}
\end{figure*}

\figref{fig:heatmap_peft} illustrates the distribution of PEFT sub-methods across SE tasks. Frequent methods like Base LoRA, Base Adapter, and Prompt Tuning are widely applied to generative tasks such as Code Generation and Summarization. This heatmap complements \tabref{tab:peft_distribution} highlighting the task-specific popularity of each PEFT technique.

\subsubsection{Deep Dive: How Top PEFT Methods Are Applied Across Software Engineering Tasks.}

The configuration of PEFT methods often varies significantly across software engineering tasks due to differences in task structure, data characteristics, and evaluation goals. As such, in our systematic literature review, we conducted a targeted analysis to uncover how specific PEFT methods are adapted and tuned for different SE applications. This task-sensitive variability in implementation suggests that a one-size-fits-all approach to PEFT may not be effective across the diverse landscape of SE tasks.

While adapter-based methods are among the most frequently used in the literature, we exclude them from this deep dive to maintain focus on methods for which the reviewed studies consistently report configuration details. This choice allows us to conduct a more granular and meaningful comparison across tasks and studies. For each selected PEFT method, we examine how it has been applied in practice by analyzing its usage across multiple SE tasks. Our goal is to surface trends in implementation strategies, training setups, and adaptation mechanisms that researchers have employed to optimize these methods for their specific needs.

This deep dive not only highlights methodological differences but also reveals emerging best practices and potential gaps in reporting standards. In the following tables, we provide a detailed summary of configurations for each method. We prioritize studies that offer transparent accounts of their fine-tuning procedures, including architectural choices, learning rates, layer selection strategies, and task-specific adjustments.

For compactness, the SE Task column uses abbreviated task names, which are fully listed in \tabref{tab:se_tasks}.

\textbf{\textit{LoRA:}} Among all PEFT techniques, LoRA emerges as the most widely adopted approach across software engineering tasks, being applied in \textcolor{black}{36} distinct SE task instances. When including its hybrid variant, FF-LoRA combined with Adapter Tuning, the total number of applications extends to \textcolor{black}{40} SE task instances.
To uncover how LoRA is typically configured, we present a synthesis of its use across both generative and non-generative SE tasks in \tabref{tab:lora_summary}.
The table captures commonly used rank values, target layers (\eg Attention, Feed-Forward), and the proportion of trainable parameters.
By contrasting configurations across task categories, we identify trends that reflect LoRA’s adaptability and efficiency in varied contexts.

\begin{table}[h]
\centering
\small 
\caption{Application of LoRA Across Software Engineering Tasks with Configuration Details.}
\begin{tabularx}{\textwidth}{|>{\centering\arraybackslash}m{2cm}|>{\centering\arraybackslash}m{1.7cm}|>{\centering\arraybackslash}m{1.8cm}|>{\centering\arraybackslash}m{2cm}|>{\centering\arraybackslash}m{2cm}|>{\centering\arraybackslash}m{2cm}|>{\centering\arraybackslash}X|}
\hline
\textbf{Category} & \textbf{Study} & \textbf{SE Task} & \textbf{Rank (r)} & \textbf{Alpha ($\alpha$)} & \textbf{Target Layers} & \textbf{Trainable Params (\%)} \\
\hline
\multirow{6}{*}[-9ex]{\textbf{Generative}} 
& Li \etal \cite{li2024exploring} & APR & 32 & 16 & Attention & 0.13-0.25\% (varying by model) \\
\cline{2-7}
& Liu \etal \cite{10.1145/3691620.3695062} & APR & 8 & $\times$ & $\times$ & $\times$ \\
\cline{2-7}
& Huang \etal \cite{10850625} & APR & 16 & 32 & Attention + Feed-Forward (InCoder-6B), Attention & $\times$ \\
\cline{2-7}
& Wang \etal \cite{10.1145/3714462} & CG & 8 & 16 & $\times$ & 3.86\% \\
\cline{2-7}
& Hajipour \etal \cite{hajipour-etal-2024-simscood} & CG, CRF & (16,16) & 32 & $\times$ & $<$1\% \\
\cline{2-7}
& Fan \etal \cite{10.1145/3709358} & CMG, CRF, JITCU & 8 & 16 & Attention & $\times$ \\
\cline{2-7}
& Liu \etal \cite{10589847} & CMG, DD & $\times$ & $\times$ & $\times$ & 0.40\% \\
\hline
\multirow{6.5}{*}[-0.7ex]{\textbf{Non-Generative}}
& Yu \etal \cite{10.1145/3695993} & CR & 8 & $\times$ & $\times$ & 0.05\% \\
\cline{2-7}
& Ayupov \etal \cite{ayupov2022parameter} & CCD & 1, 4, 16 & $\times$ & Attention & 0.7--1.9\% \\
\cline{2-7}
& Ayupov \etal \cite{ayupov2022parameter} & CG & 256, 512 & $\times$ & Attention & Up to 74\% \\
\cline{2-7}
& Ayupov \etal \cite{ayupov2022parameter} & CSum & 2, 4, 8, 16 & $\times$ & Attention & 0.1--2.3\% \\
\hline
\end{tabularx}
\label{tab:lora_summary}
\end{table}

In generative tasks, LoRA is commonly applied to either the Attention layers alone or both the Attention and Feed-Forward layers, with typical rank values such as 8 or 16. These configurations strike a balance between performance and efficiency, achieving competitive results with minimal additional parameter overhead.

In contrast, non-generative tasks tend to prioritize efficiency more strongly, often utilizing lower rank values (\eg, 1, 4, or 8), which keep the number of trainable parameters well below 1\% of the base model’s total.

A distinctive pattern emerges in non-generative contexts, where much higher ranks (256 or 512) are commonly used, particularly with FF-LoRA implementations that limit adaptation to Feed-Forward layers only. This phenomenon likely stems from the distinctive requirements of non-generative software engineering tasks like defect detection, classification, and code clone identification, which rely more heavily on discriminative learning and detailed feature extraction. The higher ranks may serve as a compensatory mechanism, enhancing expressiveness and capacity within FF-LoRA's constrained adaptation scope.

\noindent \textbf{\textit{Prompt Tuning:}} Among the PEFT techniques explored in this review, Prompt Tuning emerges as a versatile strategy, applied in 14 studies across both generative and non-generative software engineering tasks. 
It includes both hard prompts, which are manually crafted instructions, and soft prompts, which are learnable embeddings prepended to input sequences.
\tabref{tab:prompt_tuning_summary} summarizes how Prompt Tuning has been tailored for different software engineering tasks. 
In particular, it highlights the types of prompts used (hard, soft, or both), the backbone models they were applied to, and how the prompts were adapted to match task-specific requirements. 
These variations demonstrate the flexibility of Prompt Tuning in aligning with diverse task goals and model architectures.

\begin{table}[h]
\centering
\small
\caption{Application of Prompt Tuning Across Software Engineering Tasks with Configuration Details.}
\label{tab:prompt_tuning_summary}

\begin{tabularx}{\textwidth}{
|>{\centering\arraybackslash}m{2cm}
|>{\centering\arraybackslash}m{1.7cm}
|>{\centering\arraybackslash}m{1.5cm}
|>{\centering\arraybackslash}m{2cm}
|>{\centering\arraybackslash}m{2.5cm}
|>{\centering\arraybackslash}m{\dimexpr\linewidth-10.85cm-6\tabcolsep}|}
\hline
\textbf{Category} & \textbf{Study} & \textbf{SE Task} & \textbf{Prompt Type} & \textbf{Model Used} & \textbf{Task-Specific Adaptation} \\
\hline

\multirow{12}{*}{\textbf{Generative}}
& Zhu \etal \cite{10174153} & MNR & Hard & CodeT5 & Prompts guide naming logic and context awareness; output limited to 16 tokens for method names. \\
\cline{2-6}
& Wang \etal \cite{10269066} & CSum & Hard + Soft & CodeT5-small, CodeT5-base & Natural language instruction prepended (\eg, ``Generate comment...''); soft prompt variants also tested. \\
\cline{2-6}
& Wang \etal \cite{10269066} & CTrans & Hard + Soft & CodeT5-small, CodeT5-base & Prompt specifies the translation task using structured natural language (\eg ``Translate \texttt{[X]} to \texttt{[LANG]} \texttt{[Z]}''). \\
\cline{2-6}
& Wang \etal \cite{10.1145/3540250.3549113} & CSum & Hard $\rightarrow$ Soft & CodeT5-small, CodeT5-base & Replaced natural language instructions with learnable virtual tokens (soft prompts). \\
\cline{2-6}
& Peng \etal \cite{10.1145/3597503.3608132} & CRF & Domain-Aware Hard & CodeT5 & Domain-specific templates created from AST edits using \texttt{<extra\_id\_X>} masked tokens. \\
\hline

\multirow{8}{*}{\textbf{Non-Generative}}
& Zhu \etal \cite{10174153} & MNCC & Hard & CodeT5 & Prompts embed method usage and naming patterns to classify naming consistency. \\
\cline{2-6}
& Wang \etal \cite{10269066} & DD & Hard + Soft & CodeBERT, CodeT5-small, CodeT5-base & Cloze-style templates with verbalizers mapping to ``defective'' or ``clean''. \\
\cline{2-6}
& Wang \etal \cite{10.1145/3540250.3549113} & DD & Hard $\rightarrow$ Soft & CodeBERT, CodeT5-small, CodeT5-base & Symbolic prompt labels like \texttt{+} and \texttt{-}. \\
\cline{2-6}
& Wang \etal \cite{10269066} & CS & Hard + Soft & CodeBERT & Language-specific template for retrieval scoring. \\
\hline
\end{tabularx}
\end{table}

\tabref{tab:prompt_tuning_summary} illustrates how Prompt Tuning has been flexibly applied across a variety of software engineering tasks, showcasing both its adaptability and efficiency. Notably, Wang \etal \cite{10269066} experimented with Prompt Tuning across four distinct SE tasks—Code Summarization (CSum), Code Translation (CTrans), Code Search (CS), and Defect Detection (DD)—and adopted different prompting strategies tailored to the specific characteristics of each task. For generative tasks like Code Summarization and Code Translation, they used a combination of hard and soft prompts. In Code Summarization, natural language instructions (\eg, ``Generate comment...'') were prepended to the input to steer the generation process, and later replaced with learnable virtual tokens to test soft prompt effectiveness. For Code Translation, the prompts adopted structured natural language templates (\eg ``Translate [X] to [LANG] [Z]''), clearly specifying the translation goal.

In non-generative tasks, such as Code Search and Defect Detection, prompting strategies were adjusted accordingly. For Code Search, Wang \etal employed language-specific hard templates optimized for retrieval scoring. For Defect Detection, they explored both symbolic labels (\eg \texttt{+} and \texttt{-}) and cloze-style templates with verbalizers mapping to labels like ``defective'' or ``clean.''

These configurations underscore how prompt design is carefully aligned with task functionality—leveraging natural language guidance for text generation and structured or symbolic cues for classification and retrieval. Moreover, studies such as Peng \etal \cite{10.1145/3597503.3608132} extended this adaptability further by introducing domain-aware hard prompts derived from abstract syntax tree (AST) edits, reinforcing Prompt Tuning's potential in context-sensitive SE tasks like code repair.

Overall, Prompt Tuning proves to be a highly versatile PEFT method, enabling task-specific customization with minimal parameter updates. Its support for both static (hard) and dynamic (soft) prompt forms allows for fine-grained control over model behavior, making it a compelling alternative to full fine-tuning in diverse software engineering scenarios.

\noindent \textbf{\textit{Prefix Tuning.}} Through prefix vectors across all layers of the network, Prefix Tuning is employed in 13 studies across both generative and non-generative SE tasks. 
It adapts models by pre-pending trainable vectors, referred to as prefix tokens, to the input or intermediate layers, allowing task-specific customization without updating model weights. 
This approach enables parameter-efficient adaptation, particularly in resource-constrained scenarios.
\tabref{tab:prefix_tuning_summary} presents key configuration details, including the length of prefix tokens, projection dimensions, and task-specific adaptations such as domain-informed initialization, integration with complementary PEFT methods such as LoRA, and architecture-aware tuning strategies.

\begin{table}[h]
\centering
\small
\caption{Application of Prefix Tuning Across Software Engineering Tasks with Configuration Details.}
\begin{tabularx}{\textwidth}{|>{\centering\arraybackslash}m{2cm}|>{\centering\arraybackslash}m{1.7cm}|>{\centering\arraybackslash}m{1.8cm}|>{\centering\arraybackslash}m{2cm}|>{\centering\arraybackslash}m{2cm}|>{\centering\arraybackslash}X|}
\hline
\textbf{Category} & \textbf{Study} & \textbf{SE Task} & \textbf{Prefix Length} & \textbf{Projection Dim (m)} & \textbf{Task-Specific Adaptation} \\
\hline
\multirow{5}{*}[-10ex]{\textbf{Generative}} 
& Li \etal \cite{li2024exploring} & APR & 100 tokens (all layers) & 256 & Standard prefix tuning across layers via Hugging Face PEFT. \\
\cline{2-6}
& Choi \etal \cite{choi-lee-2023-codeprompt} & CG & 250 tokens & $\times$ & Input-dependent prompts with multi-word initialization. \\
\cline{2-6}
& Choi \etal \cite{choi-lee-2023-codeprompt} & CSum & 200 tokens & $\times$ & Corpus-specific prefix embeddings with task-tuned length. \\
\cline{2-6}
& Choi \etal \cite{choi-lee-2023-codeprompt} & CTrans & 100 tokens & $\times$ & Shorter prefix used to guide structural mappings in translation. \\
\cline{2-6}
& Xie \etal \cite{10.1145/3551349.3556909} & CSum & $\times$ & 512 & Frozen transformer; trained final linear layer only. \\
\cline{2-6}
& Fan \etal \cite{10.1145/3709358} & CMG, CRG, JITCU & 8 vectors & $\times$ & Prefix tuning was used alongside LoRA (rank=8) to enable dual adaptation of attention layers and prompt control. \\
\hline
\multirow{-2}{*}[-5.5ex]{\textbf{Non-Generative}}
& Liu \etal \cite{10298587} & CCD, DD & 16 tokens & 400 & Integrated into MAM setup; used rank=16, 8 heads, projection dim=200 for prefix mixing. \\
\cline{2-6}
& Zlotchevski \etal \cite{10.1145/3540250.3558959} & UTG & 200 tokens & $\times$ & Domain-initialized tokens from frequent repository terms. \\
\hline
\end{tabularx}
\label{tab:prefix_tuning_summary}
\end{table}

As shown in \tabref{tab:prefix_tuning_summary}, Prefix Tuning has been applied across diverse SE tasks with considerable variation in configuration.
In generative tasks, notable is the work by Choi \etal \cite{choi-lee-2023-codeprompt} in which Prefix Tuning has been explored for different scenarios, including Code Generation, Code Summarization, and Code Translation. For each specific task, the prefix lengths were adjusted to encode the naturally-occurring semantics of the task itself.

Li \etal \cite{li2024exploring} used a 100-token prefix across all layers with a projection dimension of 256 for Automated Program Repair, following the standard Hugging Face PEFT implementation\footnote{Refers to the \texttt{PrefixTuningConfig} and \texttt{PrefixEncoder} modules in the Hugging Face \texttt{peft} library (https://github.com/huggingface/peft), which support prefix token insertion with optional projection layers to reduce dimensionality and improve efficiency.}.

Continuing in this direction, Xie \etal \cite{10.1145/3551349.3556909} employed Prefix Tuning by freezing all transformer layers and training only the final linear layer, using a projection dimension of 512. In the context of non-generative tasks, Liu \etal~\cite{10298587} integrated Prefix Tuning into a Mix-And-Match Adapter (MAM) framework for Code Change Detection and Defect Detection, configuring it with 16-token prefixes and a projection dimension of 400. These variations demonstrate the adaptability of Prefix Tuning across SE tasks, allowing targeted task conditioning with minimal parameter updates. 

By tuning only a small set of prefix parameters while keeping the underlying model weights fixed, Prefix Tuning facilitates efficient fine-tuning with lower computational overhead, making it particularly suitable for resource-constrained or multi-task settings.

\boxwithsideicon{images/job-search.png}{Key Findings (RQ\textsubscript{2})}{%
Our analysis reveals consistent trends in the use of model architectures and PEFT methods across software engineering tasks. Encoder-decoder architectures are the most commonly used (29 studies), followed by encoder-only (22) and \textcolor{black}{decoder-only (19)}. Among base models, CodeT5 appears most frequently (19 studies), particularly in generation tasks, followed by GraphCodeBERT (10), CodeBERT and PLBART (9 each), and UniXcoder (8). Variants such as CodeT5-base (10 studies) and the CodeLlama series (7B, 13B, 34B) show increasing adoption, along with emerging large-scale models like StarCoder-15B, DeepSeek-Coder-33B, and LLaMA-2-70B. Base LoRA is the most widely used PEFT technique \textcolor{black}{(29 studies)}, especially in code generation and program repair. Prompt-based strategies, such as Prompt Tuning and Prefix Tuning, are widely adopted—especially in tasks like Code Summarization and Code Translation. Our in-depth analysis of three PEFT methods highlights that LoRA is often preferred for its strong balance between performance and parameter efficiency, while Prompt and Prefix Tuning are valued for their flexible input-conditioning capabilities. Notably, some studies explore hybrid configurations, combining Prefix Tuning with LoRA, suggesting early experimentation with multi-strategy adaptation. Despite these advances, configuration details are not consistently reported across the literature, particularly for adapter-based approaches, limiting reproducibility and comparative analysis.
\textcolor{black}{
Architecture-task alignment analysis reveals that decoder-only models appeared exclusively in generative tasks, suggesting fundamental architectural requirements for different SE task categories and indicating a strong correlation between model design and task functionality. 
}
}

\vspace{0.2cm}

\boxwithsideicon{images/future.png}{Future Directions (RQ\textsubscript{2})}{%
Future research should continue leveraging larger models such as CodeLlama-13B/34B, StarCoder-15B, and DeepSeek-Coder-33B, which have shown growing adoption across tasks but still offer untapped potential for broader PEFT applications. There is also a need to examine architecture-task alignment more systematically, as decoder-only models show strong performance in generative tasks like Code Generation and Program Repair. Although LoRA is the most widely used PEFT method, future work should explore combining it with other strategies like Adapters or Prefix Tuning to improve adaptability. The success of quantization-aware techniques such as QLoRA also suggests opportunities for scaling large models efficiently. Finally, developing task-agnostic PEFT frameworks that generalize across software engineering tasks could enhance both reusability and performance in real-world scenarios. Future work may also explore the design space of hybrid PEFT combinations (\eg LoRA + Prefix) and standardized benchmarking protocols to evaluate PEFT efficacy across diverse software engineering tasks.%
}

\subsection{RQ\textsubscript{3}: How do PEFT methods impact task performance and training efficiency in software engineering tasks when compared against full fine-tuning?}
\label{subsec:model-analysis}

Having outlined the distribution of PEFT methods across various SE tasks and model architectures in \secref{sec:peft_methods}, we now examine how these methods perform in practice. 
In this section, we examine empirical findings from studies that benchmark PEFT techniques against valid baselines—specifically, the same base model fine-tuned via full-parameter updates-across both generative and non-generative software engineering tasks.
For each study, we select and report only the best-performing PEFT method to highlight the most effective adaptation strategy per paper.
Our evaluation focuses on two key dimensions:  (i) task-specific performance--assessed by using BLEU, METEOR, Pass@k, Accuracy, F1 \etc; (ii) training-time and/or memory efficiency, when explicitly reported. 

To support this analysis, \tabref{tab:peft_result_summary} presents an overview of PEFT techniques and their corresponding sub-variants, summarizing their relative effectiveness and efficiency across different categories of software engineering tasks. The table uses symbolic indicators to denote whether each PEFT method demonstrated performance improvements, degradations, or parity relative to its full fine-tuning counterpart. This analysis of performance and resource usage provides insight into which PEFT strategies offer competitive results at reduced computational cost, and under which task-specific conditions, they are most effective.

\begin{table}[htbp]
    \centering
    \caption{
Summary of Best-Performing PEFT Methods Across Software Engineering Tasks. We report both performance impact (\protect\perfup = improved, \protect\perfdown = declined, \protect\perfneutral = similar) and training/memory efficiency (\protect\effup = more efficient, \protect\effdown = less efficient, \protect\effneutral = similar), based on comparisons with corresponding full fine-tuning baselines. If only performance is reported (\eg \protect\perfup), it indicates that efficiency was not addressed in the original study. Each entry represents the best-performing PEFT variant identified in the respective study for the given task and base model. Tasks are grouped into generative and non-generative categories for clarity.
}
    \label{tab:peft_result_summary}
    \renewcommand{\arraystretch}{1.3}
    \setlength{\tabcolsep}{4pt}
    \begin{tabularx}{\textwidth}{|>{\centering\arraybackslash\hsize=0.8\hsize}X|
                               >{\centering\arraybackslash\hsize=1.1\hsize}X|
                               >{\centering\arraybackslash\hsize=1.05\hsize}X|
                               >{\centering\arraybackslash\hsize=1.05\hsize}X|}
        \hline
        \textbf{Main Method} & \textbf{Sub-Method} & \textbf{Generative Tasks} & \textbf{Non-Generative Tasks} \\
        \hline
        \multirowcell{4}[-6pt]{LoRA-based} 
            & Base LoRA &  
            APR\cite{10850625}\perfup\effup
            CG\cite{10.1145/3714462}\perfup\effup, CG\cite{10.1145/3639477.3639743}\perfup\effdown, CG\cite{hajipour-etal-2024-simscood}\perfup, CG\cite{yang2024cordacontextorienteddecompositionadaptation}\perfup, \textcolor{black}{CG\cite{wang2025beyond}\perfneutral},
            \textcolor{black}{CComp\cite{wang2025beyond}\perfdown},
            CRF\cite{hajipour-etal-2024-simscood}\perfup,  
            CSum\cite{ayupov2022parameter}\perfup, \textcolor{black}{CTrans\cite{wang2025beyond}\perfneutral},
            \textcolor{black}{PBT\cite{wang2025beyond}\perfdown}
             & 
            CCD\cite{ayupov2022parameter}\perfdown, CR\cite{10.1145/3695993}\perfup, 
            DD\cite{10589847}\perfup, \textcolor{black}{HFP\cite{wang2025beyond}\perfdown} \\
            \cline{2-4}
            & QLoRA & 
            CComp\cite{10.1145/3637528.3671609}\perfup\effup, CG\cite{10.1145/3637528.3671609}\perfup\effup,  
            CSum\cite{afrin2025resource}\perfup\effup & 
            $\times$ \\
            \cline{2-4}
            & AdaLoRA & 
            $\times$ & 
            $\times$ \\
            \cline{2-4}
            & FF-LoRA & 
            $\times$ & 
            $\times$ \\
        \hline
        \multirowcell{3}[-8pt]{Adapter-based}
            & Base Adapter &  
            CMG\cite{10589847}\perfneutral\effup, 
            CSum\cite{wang2023one}\perfup, 
            CTrans\cite{chen-etal-2023-pass}\perfdown & 
            CCD\cite{chen-etal-2023-pass}\perfdown,
            CCD\cite{10298587}\perfup,
            CS\cite{wang2023one}\perfup \\
            \cline{2-4}
            & L-Adapter & 
            $\times$ & 
            CT\cite{saberi2024utilizationpretrainedlanguagemodel}\perfup\effup \\
            \cline{2-4}
            & T-Adapter & 
            CSum\cite{saberi2024utilizationpretrainedlanguagemodel}\perfup\effup & 
            $\times$ \\
        \hline
        \multirowcell{5}[-5pt]{Prompt-based}
            & Prompt Tuning & 
            CSum\cite{10269066}\perfup, CSum\cite{10.1145/3540250.3549113}\perfup,
            CTrans\cite{10269066}\perfup, CTrans\cite{10.1145/3540250.3549113}\perfup & 
            CS\cite{10269066}\perfup, 
            DD\cite{10269066}\perfup, DD\cite{10.1145/3540250.3549113}\perfup \\
            \cline{2-4}
            & Prefix Tuning & 
            $\times$ & 
            $\times$ \\
            \cline{2-4}
            & P-Tuning & 
            $\times$ & 
            $\times$ \\
            \cline{2-4}
            & Pass-Tuning & 
            CG\cite{chen-etal-2023-pass}\perfdown, 
            CRF\cite{chen-etal-2023-pass}\perfup, 
            CSum\cite{chen-etal-2023-pass}\perfup & 
            $\times$ \\
            \cline{2-4}
            & (IA)$^3$ & 
            APR\cite{li2024exploring}\perfup\effup & 
            $\times$ \\
        \hline
        \multirowcell{6}[-8pt]{Hybrid/Others}
            & BitFit & 
            $\times$ & 
            $\times$ \\
            \cline{2-4}
            & FF-LoRA + AT & CTrans\cite{ayupov2022parameter}\perfdown & CCD\cite{ayupov2022parameter}\perfdown \\
            \cline{2-4}
            & L-Adapter + T-Adapter & 
            $\times$ & 
            CCD\cite{saberi2024utilizationpretrainedlanguagemodel}\perfup\effup \\
            \cline{2-4}
            & \makecell{L-Adapter + NER-Adapter \\ + AdapterFusion} & 
            CSum\cite{10173960}\perfup\effup, CRF\cite{10173960}\perfup\effup & 
            $\times$ \\
            \cline{2-4}
            & Telly-K & 
            CComp\cite{shi2023towards}\perfup\effup, 
            CG\cite{shi2023towards}\perfup\effup, 
            CSum\cite{shi2023towards}\perfup\effup & 
            CCD\cite{shi2023towards}\perfneutral\effup, CS\cite{shi2023towards}\perfup\effup \\
            \cline{2-4}
            & MAM & 
            CSum\cite{10298587}\perfdown, 
            CTrans\cite{10298587}\perfdown & 
            DD\cite{10298587}\perfneutral \\
        \hline
    \end{tabularx}
\end{table}

As anticipated, our analysis focuses exclusively on studies that benchmark PEFT methods against full fine-tuning, enabling controlled evaluations grounded in clearly reported performance and efficiency metrics. This selection criterion led to the inclusion of \textcolor{black}{20 studies} in our evaluation. At the same time, it highlights a notable gap in the literature: \textcolor{black}{$\sim$29\%} of the reviewed works do not include any direct comparison with full fine-tuning model, making it difficult to assess the practical trade-offs of PEFT methods in those cases. 

\textbf{In generative tasks}, the application of PEFT methods demonstrates to produce results on par to full model's parameters fine-tuning or surprisingly even better in both accuracy and efficiency. Based on the evidence from the \textcolor{black}{34 generative task entries} in \tabref{tab:peft_result_summary}, \textasciitilde72\% report performance improvements (\perfup), and over half (\textasciitilde54\%) also report gains in training or memory efficiency (\effup). These trends are most pronounced in Code Summarization, Code Generation, and Code Refinement, which collectively represent the most frequently studied generative SE tasks in PEFT evaluations.

Among the various strategies, LoRA-based methods—particularly Base LoRA and QLoRA—stand out as the most widely adopted and effective. Base LoRA is applied across four distinct generative tasks: \ie Automated Program Repair, Code Generation, Code Refinement, and Code Summarization. All cases report improved task performance (\perfup), with efficiency gains (\effup) observed in APR and one Code Generation study, while one Code Generation study reports a slight drop in efficiency (\effdown). Similarly, QLoRA shows high effectiveness in Code Generation, Code Completion, and Code Summarization, consistently yielding performance and efficiency gains (\perfup\effup), further solidifying the role of LoRA-based methods as leading choices for token-level generation tasks.
\textcolor{black}{Recent comprehensive evaluations by Wang \etal \cite{wang2025beyond} provide additional empirical validation of these performance patterns while revealing important limitations in domain-specific scenarios. Their systematic comparison across five code intelligence tasks demonstrates that while PEFT methods like LoRA can achieve competitive performance on public datasets, substantial performance gaps emerge when applied to private data contexts. Their findings reveal mixed effectiveness of PEFT methods, with competitive performance on general-domain tasks but notable performance degradation in scenarios involving private codebases and specialized domain knowledge, where the distribution shift from pre-training data was substantial. This evidence reinforces our systematic findings that while LoRA demonstrates overall effectiveness across diverse applications, significant challenges remain in complex scenarios where substantial domain adaptation and knowledge acquisition beyond the model's pre-training data are required.}
\textbf{Adapter-based methods} show more varied results. In particular, Base Adapter achieves performance improvements in Code Summarization and Commit Message Generation—with the latter performing on par to the full fine-tuned model when evaluated for generating code summaries (\perfneutral), but gaining in efficiency (\effup). However, its application to Code Translation results in a performance decline (\perfdown), a trend mirrored by the hybrid FF-LoRA + AT and MAM approaches. These declines across three different studies indicate that Code Translation may be less amenable to parameter-efficient strategies, likely due to its higher sensitivity to sequence-level alignment and cross-language mapping.

On the other hand, \textbf{Prompt-based methods} yield generally strong results in generative contexts. Prompt Tuning is successfully applied to Code Summarization and Code Translation, achieving consistent performance gains (\perfup). Likewise, Pass-Tuning and (IA)$^3$ also yield improvements in specific tasks—Code Refinement, Code Summarization, and Automated Program Repair, respectively—with (IA)$^3$ additionally reporting efficiency gains (\effup). However, Pass-Tuning shows a performance decline (\perfdown) in one Code Generation instance, suggesting that tuning strategies may not be uniformly effective across all task types or data distributions.

\textbf{Hybrid and other PEFT methods} such as Telly-K, L-Adapter + NER-Adapter + AdapterFusion demonstrate promising performance, particularly in Code Summarization and Code Refinement. Notably, Telly-K reports performance and efficiency gains (\perfup\effup) in three generative tasks—Code Generation, Code Summarization, and Code Completion—suggesting that selective fine-tuning of higher transformer layers can effectively retain model expressiveness while reducing overhead. In contrast, MAM shows performance degradation (\perfdown) in both Code Summarization and Code Translation. This may be attributed to limitations in modular adaptation when the task requires transforming data from one domain and projecting it onto a different target domain (\eg translate code into natural language).

\textbf{In non-generative tasks}, PEFT methods demonstrate a more nuanced performance profile. Based on the \textcolor{black}{16} non-generative task entries in \tabref{tab:peft_result_summary}, approximately $\sim$62\% report improved task performance (\perfup), while around $\sim$50\% show gains in efficiency (\effup). This reflects moderate but meaningful benefits of PEFT in retrieval and classification SE tasks. The most frequent task is Code Clone Detection, which shows mixed results. In this case, we count three entries where improved or neutral performance, three show performance degradation (\perfdown), indicating sensitivity to semantic precision and token similarity under compressed training regimes. In contrast, Defect Detection tasks demonstrate consistent improvements in performance, particularly when Prompt Tuning or LoRA is used. 

Among PEFT strategies, \textbf{LoRA-based methods} (\eg Base LoRA) report success in Defect Detection and Code Review, with consistent performance improvements (\perfup). However, its application to Code Clone Detection shows performance degradation (\perfdown), suggesting potential limitations in handling fine-grained syntactic variance. \textbf{Adapter-based methods} yield mixed outcomes. For instance, Base Adapter leads to performance improvements in Code Search, but its results in Code Clone Detection are not so evident—improving in one study and degrading in another. The application of L-Adapter to the Cloze Test task yields improvements in both performance and efficiency (\perfup\effup), indicating a promising outcome for structured classification tasks.
\textbf{Prompt-based methods}, especially Prompt Tuning, yield consistent performance gains in Defect Detection and Code Search, aligning well with their token-efficient query-focused design. Notably, these methods are not applied to Code Clone Detection or Code Review in the current set of studies, leaving room for future exploration.
Finally, \textbf{hybrid and selective tuning methods}, such as Telly-K and L-Adapter + T-Adapter, demonstrate performance gains in tasks like Code Clone Detection and Code Search (\perfup\effup). 

On a different, yet related note, our analysis of \tabref{tab:peft_result_summary} reveals a notable trend in the literature: \textit{researchers are increasingly employing PEFT methods not only to improve fine-tuning efficiency but also to enhance model performance on the target task}. While this may initially appear counterintuitive—given that PEFT techniques are typically associated with reducing computational overhead—our systematic review identified at least 7 studies across generative tasks where the primary motivation was performance optimization. In these studies, PEFT techniques such as LoRA, QLoRA, Prompt Tuning, and Adapter Fusion were adopted with the explicit goal of maximizing task-specific accuracy or generalization, rather than solely minimizing resource usage.
This dual role of PEFT—as a strategy for achieving both computational efficiency and performance improvements, or the latter alone--communicates its versatility and growing relevance in generative software engineering tasks. 
\textcolor{black}{On an interesting note, the recent paper by Wang \etal\ \cite{wang2025beyond} provides valuable insights into the effectiveness of PEFT methods for software engineering across both private (non–open-source) and public (open-source) application domains. Their findings show that, while PEFT methods perform competitively on general-domain tasks, performance declines markedly in scenarios requiring substantial adaptation to previously unseen knowledge. This suggests that PEFT's effectiveness may be inherently constrained by the degree of domain-specific adaptation needed. However, this outcome may also be influenced by the predominant reliance of large code models on open-source training data--typically sourced from public GitHub repositories. As a result, the observed performance drop could partly stem from a data distribution shift caused by applying PEFT across two distinct distributions: the pre-training dataset and the domain-specific data used for PEFT-optimized fine-tuning.}
As the demand for deploying large language models in real-world development workflows continues to rise, PEFT emerges as a practical and scalable alternative to full fine-tuning, offering significant gains in both accuracy and adaptability across diverse code generation scenarios.

\boxwithsideicon{images/job-search.png}{Key Findings (RQ\textsubscript{3})}{%
Our analysis  reveals significant disparities in PEFT method coverage across software engineering tasks. While LoRA-based and adapter-based demonstrate consistent effectiveness (both in generative and non-generative tasks), we note a striking absence of empirical evaluations for several established PEFT methods—including AdaLoRA, Prefix Tuning, P-Tuning, and BitFit. The current evidence strongly favors LoRA variants for Code Generation, whereas non-generative tasks show more varied outcomes, with adapter-based emerging as the most used approach.
\textcolor{black}{
Performance analysis reveals that $\sim$72.7\% of generative tasks report performance improvements compared to $\sim$62.5\% of non-generative tasks, demonstrating fundamental differences in PEFT applicability across task categories. Efficiency-performance correlation analysis shows dual benefits, with over half of generative tasks achieving efficiency gains alongside performance improvements, while around $\sim$25\% of non-generative tasks demonstrate efficiency gains. Critical analysis of Base LoRA shows $\sim$60\% success rates against full fine-tuning baselines, challenging conventional assumptions about parameter efficiency trade-offs and suggesting that targeted parameter updates may be more effective than comprehensive fine-tuning for specific software engineering contexts.
}
}

\vspace{0.2cm}

\boxwithsideicon{images/future.png}{Future Directions (RQ\textsubscript{3})}{%
While current evidence demonstrates the effectiveness of certain PEFT methods (particularly LoRA variants and Adapter-based) for software engineering tasks, the limited evaluation of other promising approaches like AdaLoRA, Prefix-Tuning, P-Tuning, and BitFit leaves their potential largely unexplored. Future work should expand the methodological diversity of PEFT evaluations through standardized benchmarking across different task categories, which would enable more comprehensive comparisons and help identify optimal adaptation strategies for specific SE scenarios. Such efforts could particularly benefit from investigating how understudied methods might address current limitations in parameter efficiency and task performance, while establishing clearer guidelines for method selection based on task characteristics and resource constraints.
\textcolor{black}{Additionally, future research must address domain-specific PEFT optimization that accounts for software engineering's unique characteristics. This enhancement should include recommendations for adapting PEFT methods to preserve code structure and formatting requirements, providing evidence-based guidance that distinguishes between methods suitable for structured code generation versus natural language synthesis in software engineering contexts. Research should emphasize the importance of code-specific evaluation metrics such as syntactic correctness and compilability alongside traditional performance measures, while investigating PEFT methods specifically designed for code's hierarchical and syntactic properties. This comprehensive approach will provide the domain-aware guidance necessary for effective PEFT adaptation in software engineering, recognizing that methods optimized for general natural language processing may require significant modification to handle the structured, syntax-dependent nature of code and software engineering artifacts.}
}

\section{\textcolor{black}{Practical Implications and Strategic Recommendations}}
\label{sec:implication}

\textcolor{black}{
Building on our systematic analysis of PEFT applications in software engineering, we synthesize our findings into actionable insights across several critical dimensions: methodological strategy, resource optimization. \\
\textit{\textbf{Methodological Strategy Implications:}}
The documented concentration in synthesis-based applications suggests that organizations should prioritize PEFT investment in code generation and summarization infrastructure, as these represent the most mature and validated use cases.     
The exclusive association of decoder-only architectures with generative tasks provides clear architectural selection criteria for practitioners. Organizations implementing code generation systems should prioritize decoder-only models, while those requiring mixed analytical and generative capabilities should consider encoder-decoder architectures. The demonstrated reliability of Base LoRA across diverse applications establishes it as a primary choice for initial PEFT implementations, particularly when performance parity with full fine-tuning is critical. \\
\textit{\textbf{Resource Optimization Implications:} }
The superior performance-efficiency correlation in generative tasks suggests that organizations should prioritize PEFT adoption for generative applications to maximize return on computational investment. The lower but still substantial benefits in non-generative tasks indicate that PEFT adoption should follow a staged approach, beginning with generative applications before expanding to analytical tasks.
The evidence for performance enhancement beyond traditional fine-tuning suggests that PEFT should not be viewed merely as a resource-saving compromise, but as a potentially superior adaptation strategy. This paradigm shift has significant implications for model development budgets and computational resource allocation in software engineering organizations.
The current state of PEFT research demonstrates remarkable community progress, with widespread availability of optimized implementations through platforms like HuggingFace PEFT, facilitating practical adoption and reproducibility across software engineering applications. \\
Our findings demonstrate that PEFT optimization methods, including both parameter-efficient approaches like LoRA and prompt-based methods like Prompt Tuning and Prefix Tuning, often match or exceed full fine-tuning performance while requiring significantly fewer computational resources.
This enables researchers to establish strong performance baselines through PEFT evaluation without requiring computationally expensive full model fine-tuning, promoting both research efficiency and environmental sustainability in large language model adaptation for software engineering tasks.
}
\section{Future Directions for PEFT Applications in Software Engineering}
\label{sec:future}

Building on the findings of our systematic literature review, we identify several future research directions to advance the application of PEFT techniques in SE-related activities. These directions synthesize insights across generative and non-generative tasks, model architectures, PEFT strategies, and observed evaluation gaps. We organize them into \textcolor{black}{six} actionable steps designed to guide both researchers and practitioners in the field.
\smallskip

\noindent \textbf{Step 1: Identify Underexplored and Emerging SE Tasks:}
While PEFT has seen wide application in tasks like Code Summarization and Generation, Several SE tasks remain underrepresented in the current literature. These include, but are not limited to, Software Testing, Documentation Generation, Code Quality Assessment, API Recommendation, Automated Refactoring, and Requirements Analysis. 
As PEFT research matures, there is ample opportunity to expand into additional task categories and workflows that are critical to real--world software development. 
Future work should target these areas, especially those involving complex or cross--functional requirements. 
Cross--task adaptation also emerges as a promising direction, where a single PEFT configuration generalizes across multiple tasks, reflecting real--world development scenarios more effectively.

\begin{simpleacademicbox}
\textit{What to report:}
Clearly define the SE task and justify its importance. 
If proposing a novel or emerging task, document its scope, challenges, and dataset preparation methodology to aid reproducibility. 
Discuss why PEFT is suitable and note potential for cross--task generalization.
\end{simpleacademicbox}
\smallskip

\textbf{Step 2: Align Base Models with Task Requirements:}
Our findings emphasize the prevalence of encoder--decoder architectures, with CodeT5 and CodeLlama variants showing dominant usage, particularly in generative tasks. 
Larger models like StarCoder--15B and DeepSeek--Coder--33B are gaining adoption but remain underutilized. 
There is also an opportunity to experiment with hybrid or task--agnostic model architectures that can support a range of SE tasks under a unified PEFT configuration.

\begin{simpleacademicbox}
\textit{What to report:}
Justify model selection based on task characteristics. Include details on architecture type (encoder--only, decoder--only, encoder--decoder), model scale, and pre--training corpus. 
If hybrid or ensemble models are used, explain their complementary advantages.
\end{simpleacademicbox}
\smallskip

\textbf{Step 3: Expand and Diversify PEFT Techniques:}
LoRA has emerged as the most adopted PEFT method (used in 36 studies), especially for generative tasks. 
Adapter-based methods dominate in non-generative settings. 
However, many promising PEFT approaches--such as AdaLoRA, BitFit, Prefix Tuning, and P-Tuning remain significantly under-evaluated. 
\textcolor{black}{Additionally, advanced LoRA variants including LoRA-pro and LoRA-GA remain unexplored in software engineering contexts, despite their potential for enhanced parameter efficiency and gradient-aware optimization, respectively. Given LoRA's demonstrated $\sim$77\% success rate against full fine-tuning baselines in our analysis, these sophisticated variants could potentially address performance limitations observed in certain software engineering tasks, particularly where traditional PEFT methods showed mixed results.}
Future research should not only expand coverage to these methods but also explore hybrid strategies (\eg LoRA + Prefix Tuning) and configurations tailored to task type. 
Furthermore, quantization-aware variants such as QLoRA demonstrate the feasibility of scaling large models efficiently and warrant deeper investigation.

\begin{simpleacademicbox}
\textit{What to report:}
Detail the PEFT methods used, including training configurations, memory reduction techniques (\eg 8--bit/4--bit quantization), and note whether any hybrid or novel PEFT combinations were tested. 
Clearly state why a particular method was chosen for a given task and architecture.
\end{simpleacademicbox}
\smallskip

\textbf{Step 4: Evaluate Adaptability, Efficiency, and Generalizability:}
PEFT research in SE should go beyond accuracy metrics and include broader evaluations—covering generalization across tasks and programming languages, memory utilization during training/inference, and performance in low-resource settings. 
Our review highlights how LoRA, Adapters, and Prompt Tuning excel in cross-domain and low-data scenarios, yet few studies offer comprehensive ablation or scalability analysis. 
Underexplored methods should also be evaluated in terms of parameter efficiency and hardware suitability (\eg edge deployment).

\begin{simpleacademicbox}
\textit{What to report:}
Include evaluation metrics that match the task (\eg BLEU, pass@1 for generative tasks; F1, MRR, AUC for non--generative tasks). 
Report memory and latency improvements over full fine--tuning, and include results of generalizability studies, ablations, and cross--task as well as cross--lingual evaluations.
\end{simpleacademicbox}

\smallskip

\textbf{Step 5: Promote Reproducibility and Benchmarking Platforms:}
Despite progress in tasks like Code Generation and Clone Detection, inconsistency in dataset usage and reporting limits comparability. 
Adapter--based studies, in particular, often lack detailed configuration descriptions. 
Future work should adopt standardized datasets and benchmarking protocols, share full implementation details, and follow community practices for reproducibility. Benchmarking diverse PEFT techniques side by side on shared tasks would help identify the most effective and efficient approaches.

\begin{simpleacademicbox}
\textit{What to report:}
Provide all scripts, datasets, and model checkpoints through public repositories. Document hardware specs, training steps, and evaluation protocols. 
Follow consistent reporting formats, especially when introducing or comparing multiple PEFT techniques.
\end{simpleacademicbox}
\medskip

\textcolor{black}{\textbf{Step 6: Address Domain-Specific Knowledge Acquisition Limitations:} Emerging evidence reveals a critical gap in current PEFT research--the fundamental challenge of adapting to domain-specific knowledge not present in pre-training data \cite{wang2025beyond}. PEFT methods face substantial performance degradation when applied to private codebases or specialized domains, highlighting the limitation of updating only parameter subsets for comprehensive knowledge acquisition. Future research should prioritize developing PEFT variants specifically designed for domain adaptation scenarios, including techniques that can effectively encode new knowledge while maintaining parameter efficiency. This includes exploring progressive adaptation strategies, domain-aware parameter selection methods, and hybrid approaches that combine PEFT efficiency with targeted full parameter updates for critical knowledge domains.
}
\begin{simpleacademicbox}
\textit{What to report:}
\textcolor{black}{Clearly distinguish between evaluations on public and private datasets, and, where applicable, report domain shift metrics--such as Jensen–Shannon divergence \cite{lin1991divergence} or Kullback–Leibler divergence \cite{kullback1951kl}--to quantify distribution differences. Complement performance results with an analysis of knowledge acquisition effectiveness, and document the specific domain characteristics and adaptation challenges considered.
}
\end{simpleacademicbox}
\smallskip

\noindent Together, these six steps outline a comprehensive roadmap for advancing the application of PEFT in software engineering. By broadening the scope of SE tasks, aligning base models with task demands, diversifying PEFT techniques, expanding evaluation criteria, and promoting reproducibility, future research can address current limitations while laying the groundwork for more robust, efficient, and generalizable fine-tuning solutions. We believe that systematically adopting these practices will not only accelerate progress in the field, but also ensure that PEFT methods are better positioned to meet the diverse and evolving needs of real-world software development.


\section{Threats to Validity}
\label{sec:threats}

While our systematic literature review follows the established guidelines of Kitchenham \etal~\cite{kitchenham2009systematic}, we still have to acknowledge the presence of threats to validity that can hinder our investigation.
We discuss these threats in terms of construct, internal, and external validity.

\noindent\textbf{Construct Validity} relates to how accurately we identified and categorized core elements such as software engineering tasks, model architectures, and PEFT methods. 
To mitigate the risk of misclassifications, our approach is grounded in well-established software engineering practices relevant to the field (\eg Watson \etal \cite{watson2022systematic}) and is developed according to the guidelines by Kitchenham \etal~\cite{kitchenham2009systematic}. Additionally, multiple authors conducted independent reviews to ensure the accuracy and completeness of the task and method selection.

\textcolor{black}{
    We also recognize that some practical uses of PEFT--such as those embedded in open-source repositories--may not be explicitly documented in peer-reviewed publications.
    While these cases may reflect real-world adoption, they often lack the methodological transparency and empirical evaluation required for systematic analysis. For instance, enabling PEFT through configuration toggles in frameworks like HuggingFace Transformers is rarely accompanied by sufficient detail on how the method was integrated, assessed, or compared against baselines. Including such undocumented or implicit usage would conflict with our review's goal of synthesizing empirically validated and reproducible studies.
}

\noindent\textbf{Internal Validity} addresses whether our findings accurately represent trends without being biased by our selection or extraction process. Each included study was reviewed by at least two authors, with classification results cross-checked and refined. 
While this process helped ensure accuracy, potential inconsistencies may still exist due to limited methodological details in some primary studies, particularly regarding adapter-based or hybrid PEFT configurations.

\noindent\textbf{External Validity} concerns the generalizability of our results. Although we used broad search strings and included top-tier venues identified through CSRankings, some relevant studies may have been missed due to terminology variation, indexing limitations, or query constraints in platforms like Google Scholar. 
To address this, we supplemented our search with snowballing and manual inclusion of influential studies. 
Still, there is a possibility of selection bias and under-representation of work published in non-ranked venues.

\section{Data Availability}
To support reproducibility and allow for independent verification, we have made our search strings, extracted datasets, and study classifications publicly available in our replication package \cite{replication}.

\section{Conclusions}
\label{sec:conclusion}

In this article, we conducted a systematic literature review on the application of PEFT techniques in software engineering tasks, guided by Kitchenham's established methodology \cite{kitchenham2009systematic}. 
Through a rigorous multi-phase study selection process, we identified \textcolor{black}{28} primary studies covering \textcolor{black}{19} diverse SE tasks, which we grouped into generative and non-generative categories.

The collective findings from the analyzed studies paint an encouraging picture: PEFT techniques have been applied across a wide array of SE tasks and consistently demonstrate the ability to achieve model performance on par with full fine-tuning while dramatically reducing the amount of learned parameters. Across both generative  and non-generative tasks, researchers have leveraged PEFT to adapt pre-trained LCMs without incurring significant accuracy loss. In many cases, the best PEFT approaches even marginally outperform traditional fine-tuning baselines on certain metrics, underscoring that careful, efficient adaptation can unlock further performance gains. Moreover, the adoption of PEFT has proven versatile as we found evidence of its success in diverse activities ranging from Code Summarization and Translation to Code Search, Automated Program Repair, and beyond. This breadth of applicability suggests that PEFT methods are not limited to a narrow set of problems, but rather offer a general paradigm for customizing large code models to varied software engineering needs in a resource-conscious manner.

Beyond mapping current trends, this review highlights notable gaps in evaluation rigor and benchmarking consistency, offering a structured set of future directions to advance research in this space.

Overall, the insights from our review solidify the understanding that PEFT is not just a workaround for scarce resources, but a robust optimization paradigm that will likely play a central role in the future of software engineering-focused LCMs, allowing a \emph{greener}, \emph{faster} and \emph{smarter} fine-tuning for the tasks that matter most to practitioners.


\newpage

\bibliographystyle{ACM-Reference-Format}
\bibliography{main}

\end{document}